\documentclass[
 reprint,
 superscriptaddress,
 amsmath,amssymb,
 aps,
 pra,
 letter
]{revtex4-2}

\usepackage{graphicx}
\usepackage{dcolumn}
\usepackage{bm}
\usepackage{physics}
\usepackage{braket}
\usepackage{amsmath,amsfonts,mathtools,latexsym}
\usepackage{amssymb}
\usepackage{amsbsy}
\usepackage{here}
\usepackage{color}
\usepackage[percent]{overpic}
\usepackage{docmute}

\newcommand{\me}{\mathrm{e}}
\newcommand{\mi}{\mathrm{i}}
\newcommand{\mS}{\mathrm{S}}
\newcommand{\Lap}{\mathrm{Lap}}
\newcommand{\mP}{\mathcal{P}}
\newcommand{\mQ}{\mathcal{Q}}

\newcounter{panel}
\renewcommand{\thepanel}{\alph{panel}}
\newcommand{\paneltag}[1]{\refstepcounter{panel}\label{#1}(\thepanel)}

\begin{document}


\title{Fractional power-law decay in the spontaneous emission of a two-level system}

\author{Hiroki Nakabayashi}
\email{nakaba@iis.u-tokyo.ac.jp}
\affiliation{Department of Physics, The University of Tokyo, 5-1-5 Kashiwanoha, Kashiwa, Chiba 277-8574, Japan}
\author{Hayato Kinkawa}
\email{kinkawa@iis.u-tokyo.ac.jp}
\affiliation{Department of Physics, The University of Tokyo, 5-1-5 Kashiwanoha, Kashiwa, Chiba 277-8574, Japan}
\author{Takano Taira}
\email{taira.takano.292@m.kyushu-u.ac.jp}
\affiliation{Department of Physics, Kyushu University, 744 Motooka, Nishiku, Fukuoka 819-0395, Japan}
\affiliation{Institute of Industrial Science, The University of Tokyo, 5-1-5 Kashiwanoha, Kashiwa, Chiba 277-8574, Japan}
\author{Naomichi Hatano}
\email{hatano@iis.u-tokyo.ac.jp}
\affiliation{Institute of Industrial Science, The University of Tokyo, 5-1-5 Kashiwanoha, Kashiwa, Chiba 277-8574, Japan}

\date{\today}

\begin{abstract}
    It has been shown for various systems that the decay rate of an unstable quantum system deviates from exponential behavior in short- and long-time regimes associated with memory effects. In particular, it is widely believed that in the short-time regime, the decay is quadratic, inducing the quantum Zeno effect, in which the decay is suppressed by rapidly repeated measurements. In our study, we find that when the environment of the unstable system has an energy spectrum with a lower bound but without an upper one, the decay rate in both regimes is scaled in terms of the spatial dimension and the exponent of the energy dispersion of the environment. Surprisingly, we find that in the short-time regime the decay exhibits fractional scaling, which leads to a quantum Zeno effect with a different scaling of the Zeno time.
\end{abstract}

\maketitle


\noindent\textit{Introduction.—}The spontaneous decay of a two-level system embedded in an environment is typically described in terms of an exponential decay law \cite{gamow}. However, it has been shown theoretically \cite{khalfin1958contribution,fonda1978decay, peshkin2014non,chiu1977time} and experimentally \cite{wilkinson1997experimental,rothe2006violation, crespi2019experimental} that the decay deviates from the exponential law in both short- and long-time regimes. In the short-time regime,  a non-exponential decay is essential for the quantum Zeno effect \cite{misra1977zeno,chiu1977time,itano1990quantum}. In the long-time regime, the decay follows a power-law behavior with oscillations \cite{kofman1994spontaneous}.
Notably, these deviations cannot be described by a time-independent Gorini-Kossakowski-Sudarshan-Lindblad (GKSL) equation \cite{gorini1976completely,lindblad1976generators,taira2024markovianity}.

The spontaneous decay is characterized by the survival probability
$    p(t)=|\braket{\psi(0)|\psi(t)}|^2$,
where $\ket{\psi(t)}=\me^{-\mi Ht}\ket{\psi(0)}$ with $\hbar$ set to unity.
A conventional analysis claims that the decay in the short-time regime is quadratic \cite{facchi2008quantum}, based on a simple Taylor expansion with respect to time $t$:
\begin{align}
\abs{\braket{\psi(0)|\psi(t)}}^2&\simeq 1-\qty(\braket{H^2}-\braket{H}^2)t^2,\label{taylor}
\end{align}
where $\braket{\cdot}$ represents the expectation value with respect to $\ket{\psi(0)}$. This appears to be universally valid, but when we consider an open quantum system coupled to an environment that has infinite degrees of freedom and the energy spectrum without an upper bound, the energy variance $\braket{H^2}-\braket{H}^2$ may diverge, which makes the conventional discussion inapplicable. In fact, it has been reported for an unbounded spectrum that a term linear in $t$ appears in the expansion of the survival probability \cite{taira2024markovianity,nishino2024exact}.

In this letter, we address a setup in which the energy variance $\braket{H^2}-\braket{H}^2$ indeed diverges and show that the fractional power-law decay occurs in the short- and long-time regimes. More specifically, we consider a two-level system coupled to an environment with an energy dispersion that has a lower bound but without an upper one. We show that in the short-time regime, the decay is characterized by the spatial dimension $D$ and the exponent $n$ of the dispersion relation $\omega(k) \propto |k|^n$ of the environment, leading to a decay that has a fractional exponent $1-ct^{2-D/n}$ if $0<D/n< 1$, where $c$ is a constant.
 The fractional scaling of the short-time decay can also induce the quantum Zeno effect but with a different Zeno time; we call it the anomalous quantum Zeno effect. Furthermore, by using the Feshbach non-Hermitian effective Hamiltonian \cite{feshbach1958unified,hatano2014time}, we reveal that the long-time dynamics exhibits a power-law decay of the form $t^{D/n-2}$ with oscillations if $0<D/n< 1$. Note that the dispersion relations with higher powers $n\geq 3$ have recently been realized experimentally in multi-Weyl semimetals \cite{dantas2020non,schmeltzer2023optical}, in periodic dielectric waveguides \cite{burr2015zero, wood2015degenerate}, and in coupled microwave waveguides with higher-order degenerate band edges \cite{yazdi2022sixth}.
 
 In contrast, the time-independent GKSL
 equation, which results from the assumption of semigroup in the completely positive
 trace-preserving map, yields purely exponential decay in the model of the
 spontaneous emission of a two-level system considered in this letter
 \cite{taira2024markovianity, breuer2002theory, garraway1997nonperturbative};
 see Sec.~D of Supplemental Material~\cite{supplemental_material}.\nocite{gradshteyn2014table} Therefore, the non-exponential decay discussed
 in this work signals the breakdown of the semigroup description.

\noindent\textit{Model.—} We consider a model describing a spontaneous decay of a two-level system coupled with a $D$-dimensional environment with the dispersion relation of the form $\omega(\vec{k})=\omega_0|\vec{k}|^n$. We model the two-level system and the environment by the Hilbert space $\mathcal{H}_{\mS}$ and $\mathcal{H}_{\mathrm{E}}$. The composite total system is described by $\mathcal{H}_{\rm tot}=\mathcal{H}_{\mS}\otimes\mathcal{H}_{\mathrm{E}}$. We do not distinguish between an operator $X$ on $\mathcal{H}_{\mS}$ and $X\otimes I_{\rm E}$ on $\mathcal{H}_{\rm tot}$. The total Hamiltonian is given as follows:
\begin{gather}
  H=H_0 + H_{\mathrm{int}}, ~~H_0= H_{\mathrm{S}} + H_{\mathrm{E}},\label{1}\\ 
  H_{\mathrm{S}} = \omega_{\mathrm{S}} \sigma_+ \sigma_-\otimes I_{\rm E},\\
  H_{\mathrm{E}} = I_{\rm S}\otimes \omega_0 \int_{\mathbb{R}^D}\dd {\vec{k}} |\vec{k}|^n b^\dagger(\vec{k}) b(\vec{k}),\\
  H_{\mathrm{int}} =   \int_{\mathbb{R}^D}\dd {\vec{k}} g(\vec{k}) \qty[\sigma_+\otimes b(\vec{k}) + \sigma_-\otimes b^\dagger(\vec{k})]\label{4}.
\end{gather}
Here, $I_{\rm S}$ and $I_{\rm E}$ denote the identity operators for the system and the environment, respectively, and $\mathbb{R}^D$ denotes the full $D$-dimensional $k$ space.

The Hamiltonian $H_{\mathrm{S}}$ represents a two-level system with the excitation energy $\omega_{\rm S}$. We use $\sigma_\pm$ as ladder operators, which obey $\{\sigma_-,\sigma_+\}=1$ and $\sigma_\pm^2=0$, and act as $\sigma_+|0\rangle_{\rm S}=|1\rangle_{\rm S}$, $\sigma_-|1\rangle_{\rm S}=|0\rangle_{\rm S}$.
Here, $\ket{0}_{\rm S}$ and $\ket{1}_{\rm S}$ indicate the ground and excited states of the two-level system, respectively, and $\qty{\cdot,\cdot}$ represents the anti-commutator.

The Hamiltonian $H_{\mathrm{E}}$ represents the environment with the dispersion relation $\omega_0 |\vec{k}|^n$ in $D$ spatial dimensions. The operators $b^\dagger(\vec{k})$ and $b(\vec{k})$ are the creation and annihilation operators for the particle with momentum $\vec{k}$ in the environment, which satisfy 
\begin{gather}
    \qty[b(\vec{k}),b^\dagger(\vec{k}')]= \delta(\vec{k}-\vec{k}'),~\qty[b(\vec{k}),b(\vec{k}')]=0,\\
  \ket{\vec{k}}_{\rm E}=b^\dagger(\vec{k}) \ket{\rm vac}_{\rm E},~\ket{\rm vac}_{\rm E}=b(\vec{k}) \ket{\vec{k}}_{\rm E}.
\end{gather}
Here, $\ket{\vec{k}}_{\rm E}$ denotes the one-particle state with momentum $\vec{k}$, $\ket{\rm vac}_{\rm E}$ denotes the vacuum state of the environment and $\qty[\cdot,\cdot]$ represents the commutator. The Hamiltonian $H_{\mathrm{int}}$ represents the interaction between the two-level system and the environment with a coupling term $g(\vec{k})$. Throughout this paper, we set $g(\vec{k})=g$. Notice that in our model, we omitted the virtual interaction $\sigma_+b^{\dagger}(\vec{k})$ and $\sigma_- b(\vec{k})$; this corresponds to the rotating wave approximation, which we justify in Sec.~E of Supplemental Material~\cite{supplemental_material} for our short-time analysis. The total Hamiltonian (\ref{1}) commutes with the number operator $\sigma_+ \sigma_-+ \int_{\mathbb{R}^D}\dd{\vec{k}} b^\dagger(\vec{k}) b(\vec{k})$. 

We assume that initially the two-level system is in its excited state in a vacuum environment as in $\ket{\psi(0)}:=\ket{1}_{\rm S}\otimes \ket{\rm vac}_{\rm E}$ and let it evolve as in $\ket{\psi(t)}=\me^{-\mi Ht}\ket{\psi(0)}$. The Hamiltonian ($\ref{1}$) can then be written in the form of the first quantization:
\begin{align}
    H =& \omega_{\mathrm{S}} \ketbra{\mathrm{S}}{\mathrm{S}} +\omega_0 \int_{\mathbb{R}^D}\dd{\vec{k}} |\vec{k}|^{n} \ketbra*{\vec{k}}\notag \\
    &+ g \int_{\mathbb{R}^D}\dd{\vec{k}} (\ketbra*{\mathrm{S}}{\vec{k}}  + \ketbra*{\vec{k}}{\mathrm{S}} ),\label{15}
  \end{align}
where we introduced the abbreviations
\begin{gather}
     \ket{\rm S}:=\ket{1}_{\rm S}\otimes \ket{\rm vac}_{\rm E},~  \ket{\vec{k}}:=\ket{0}_{\rm S}\otimes \ket{\vec{k}}_{\rm E}.
\end{gather}
The complete set in the single-excitation sector is given by 
\begin{gather}
     I=\ketbra{\mS} +\int_{\mathbb{R}^D}\dd{\vec{k}} \ketbra*{\vec{k}},
\end{gather}
where $I$ is the identity operator of the sector.
Then, the state at any time $t$ must be given by the form
\begin{align}
  \ket{\psi(t)}=&
    c_{\rm S}(t)\ket{\mS}+\int_{\mathbb{R}^D}\dd{\vec{k}} c_{\vec{k}}(t)\ket{\vec{k}}.
\end{align}
Below, we analyze the survival amplitude $\braket{\mS|\psi(t)}=c_{\rm S}(t)$ and the survival probability $p(t)=\abs{\braket{\mS|\psi(t)}}^2$.

\noindent\textit{Short-time limit of the survival amplitude.—}
In the short-time regime, since the Taylor expansion (\ref{taylor}) is not applicable, we instead employ the Dyson series expansion, which is in combined powers of the coupling $g$ and time $t$. 
Let us calculate $\braket{\mathrm{S}|\psi(t)}$ by expanding it in the Dyson series for $\ket{\psi(t)}$ with respect to $H_{\mathrm{int}}$:
\begin{align}
    \braket{\mS|\psi(t)}&=\me^{-\mi\omega_{\mS}t}\sum_{\ell =0}^{\infty} \qty(-\mi)^{\ell} \int_{0}^{t} \dd{t_1} \int_{0}^{t_1} \dd{t_2} \cdots \int_{0}^{t_{\ell -1}} \dd{t_{\ell}}\notag\\
    &~~~~\times \braket{\mS|H_{\mathrm{int}}^{(\rm I)}(t_1)H_{\mathrm{int}}^{(\rm I)}(t_2)\cdots H_{\mathrm{int}}^{(\rm I)}(t_{\ell})|\mS},\label{23}
\end{align}
where
    $
    H_{\mathrm{int}}^{(\rm I)}(t) := \me^{\mi H_0 t} H_{\mathrm{int}} \me^{-\mi H_0 t}
    $
    is the coupling Hamiltonian in the interaction picture.
Evaluating the integral, we obtain the series solution of $\braket{\mS|\psi(t)}$ as follows:
\begin{align}
    &\braket{\mS|\psi(t)}\notag\\
    &=\me^{-\mi \omega_{\mS}t}\sum_{\ell=0}^{\infty} \qty[-\frac{S_D\omega_0^{-\nu}\Gamma(\nu)\Gamma(1-\nu)}{n}\me^{-\mi\nu\pi/2}g^{2}t^{2-\nu}]^\ell \notag\\
    &~~~\times \frac{1}{\Gamma(2\ell-\ell\nu+1)}
    \sum_{m=0}^{\infty}\frac{(\ell-\ell\nu)_m (\mi \omega_{\mS}t)^m}{(2\ell-\ell\nu+1)_m m!} ,
\end{align}
where $(\cdot)_m:=\prod_{k=0}^{m-1}(\cdot+k)$, $(\cdot)_0=1$ denote the Pochhammer symbol, $\nu=D/n$, $S_D$ is the $D-1$ dimensional surface area of the $D$-dimensional unit sphere, i.e., $S_D={2\pi^{D/2}/\Gamma(D/2)}$, and $\Gamma(\cdot)$ represents the gamma function;
 see Sec.~A of Supplemental Material~\cite{supplemental_material}.
We note that this calculation is valid only for $0<\nu<1$ because we need to avoid divergence arising in the calculation. 

Therefore the dominant term of the survival amplitude in the regime of small $t$ is given by the term $m=0$ and $\ell=1$:
\begin{align}
    &\braket{\mS|\psi(t)}\notag\\
    &\simeq \me^{-\mi \omega_{\mS}t}\qty[1-\frac{\nu \Gamma(\nu)\pi^{D/2}\me^{-\mi \nu \pi/2}}{\Gamma(\frac{D}{2}+1)(1-\nu)(2-\nu)} 
    g^2{\omega_0}^{-\nu}t^{2-\nu}].\label{amplitude}
\end{align}
Thus, in the short-time limit $t\to 0$, the survival amplitude has a fractional exponent $2-D/n$. This Dyson series expansion, which avoids the divergence inherent in the Taylor expansion (\ref{taylor}), is in fact the series in the dimensionless combination $g^2{\omega_0}^{-\nu}t^{2-\nu}$, giving the fractional exponent. Note that this leading short-time contribution remains the same even if the counter-rotating terms $\sigma_+b(k)^{\dagger}$ and $\sigma_- b(k)$ are included because they do not contribute to the second-order term $\mathcal{O}(g^2)$ in the Dyson series expansion; see Sec.~E of Supplemental Material~\cite{supplemental_material}.

The Dyson series expansion is useful in understanding the origin of this fractional exponent as follows. Let us focus on the second-order term in the expansion (\ref{23})
\begin{align}
    \eta(t_1-t_2)&:=\braket{\mS|H_{\mathrm{int}}^{(\rm I)}(t_1)H_{\mathrm{int}}^{(\rm I)}(t_2)|\mS}\label{11}\\
    &=\me^{\mi\omega_{\mS}(t_1-t_2)}\braket{\mS|H_{\mathrm{int}}\me^{-\mi H_0 (t_1-t_2)}H_{\mathrm{int}}|\mS},
\end{align}
and examine the following term:
\begin{align}
  &~~\braket{\mS|H_{\mathrm{int}}\me^{-\mi H_0 (t_1-t_2)}H_{\mathrm{int}}|\mS}\notag\\
    &={}_{\mS}\!\bra{1}{}_{\mathrm{E}}\!\bra{\mathrm{vac}}\qty[\sigma_{+}\otimes g\int_{\mathbb{R}^D} \dd{\vec{k}}b(\vec{k})]\cdot\qty[\me^{-\mi H_{\mathrm{E}} (t_1-t_2)}] \notag\\
    &~~~~\cdot \qty[\sigma_{-}\otimes g\int_{\mathbb{R}^D} \dd{\vec{k'}}b^{\dagger}(\vec{k'})]\ket{1}_{\mS}\ket{\mathrm{vac}}_{\mathrm{E}}\\
    &=(2\pi)^D g^2 {}_{\mS}\!\bra{1}{}_{\mathrm{E}}\!\bra{\mathrm{vac}} \sigma_{+}\otimes b(\vec{x}=0)\me^{-\mi H_{\mathrm{E}} (t_1-t_2)}\notag\\
    &~~~~\sigma_{-}\otimes b^{\dagger}(\vec{x}=0)\ket{1}_{\mS}\ket{\mathrm{vac}}_{\mathrm{E}},\label{27}
\end{align}
where $b(\vec{x})$ is the Fourier transform
\begin{align}
    b (\vec{x}):= \frac{1}{\sqrt{(2\pi)^D}}\int_{\mathbb{R}^D}\dd{\vec{k}} \me^{-\mi \vec{k}\cdot \vec{x}}b (\vec{k}).\label{24}
\end{align} 

Equation~(\ref{27}) represents the following time evolution. The two-level system decays and emits the particle into the environment at time $t_1$ at $\vec{x}=0$, which evolves until time $t_2$ in the environment, and then the two-level system absorbs it at $\vec{x}=0$. In other words, the autocorrelation $\eta(t_1-t_2)$ measures the probability amplitude for energy that has left the system to return after spending time $t_1-t_2$ in the environment. A finite value of the autocorrelation indicates that the dynamics of the system is coherently memorized in the environment, which provides a physical interpretation of the non-Markovianity in terms of the breakdown of the semigroup description.

We find $\eta(t_1-t_2) \propto (t_1-t_2)^{-\nu}$; see Sec.~A.2 of Supplemental Material~\cite{supplemental_material}. Then, integrating $(t_1-t_2)^{-\nu}$ over $t_2$ and $t_1$ yields the fractional exponent $2-\nu$ as in Eq.~(\ref{amplitude}). The collapse of the memory kernel $\eta(t_1-t_2)\propto\delta (t_1-t_2)$ would yield a Markovian exponential decay, e.g., for massless Dirac particles\cite{taira2024markovianity}, which could imply that no amplitude flows back to the system from the environment.

\noindent\textit{Anomalous quantum Zeno effect.—}The fractional exponent $2-D/n$ of $\braket{\mathrm{S}|\psi(t)}$ leads to a nontrivial result of the survival probability $p(t)=\abs{\braket{\mathrm{S}|\psi(t)}}^2$; see Fig.~1(b). It decays with the fractional exponent $2-\nu$ as in
\begin{gather}
  p(t)\simeq 1-2 \frac{\nu\Gamma(\nu)\pi^{D/2}\cos(\nu \pi /2)}{\Gamma(\frac{D}{2}+1)(1-\nu)(2-\nu)}g^2\omega_0^{-\nu}t^{2-\nu}\label{30}.
\end{gather}
\begin{figure}[t]
    \centering
    \begin{minipage}{0.45\columnwidth}
      \includegraphics[width=\linewidth]{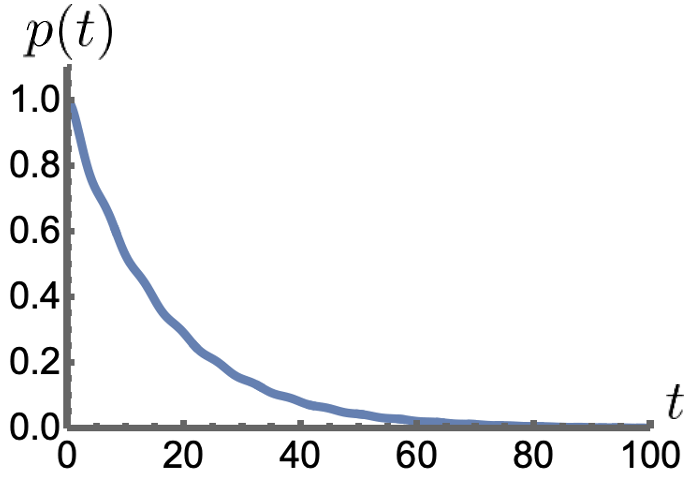}
      \par\small\centering \paneltag{fig:pt-a} 
    \end{minipage}\hfill
    \begin{minipage}{0.50\columnwidth}
      \includegraphics[width=\linewidth]{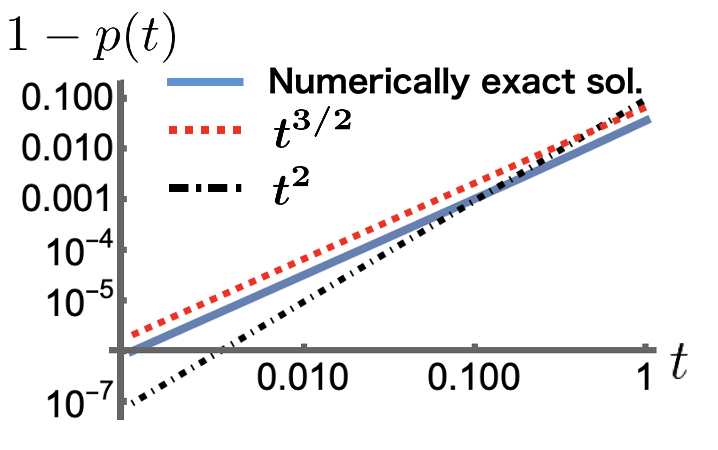}
      \par\small\centering \paneltag{fig:pt-b} 
    \end{minipage}  
    \caption{Time dependence of the survival probability $p(t)$ for $\omega_{\mathrm{S}}=1$, $g=0.1$, $D=1$ and $n=2$, and hence $2-\nu=3/2$: (a) a linear plot of $p(t)$; (b) a logarithmic plot of $1-p(t)$ along with $t^2$ and $t^{3/2}$. The numerical result is obtained by evaluating Eq.~(\ref{33}), as introduced below. Note that $p(t)$ does not vanish in the limit $t\to \infty$; the remnant is about $3.86\times 10^{-6}$, and hence invisible in Panel (a).}
  \end{figure}
The short-time behavior of the survival probability in Eq.~(\ref{30}) induces the quantum Zeno effect under rapidly repeated measurements, while its Zeno time exhibits a different scaling compared to the conventional case. 

The Zeno time \cite{petrosky2002quantum} $\tau_{\mathrm{Z}}$ associated with the quadratic decay $p(t)\simeq 1-ct^2$ is defined as $1\sim c\tau_{\mathrm{Z}}^2$, characterizing a time scale of the concave decay. It therefore scales as
$
  \tau_{\mathrm{Z}}\sim c^{-1/2}
$. 
In our case, in which the short-time behavior exhibits the decay with a fractional exponent $p(t) \simeq 1-ct^{2-\nu}$, the Zeno time scales as
$
  \tau_{\mathrm{Z}}\sim c^{-{1}/{(2-\nu)}}.
$
This implies that by adjusting the parameter $\nu=D/n$, we can control the characteristic time scale of the quantum Zeno effect. When $c<1$, i.e. for a weak coupling, the Zeno time associated with the fractional power-law decay is longer than that of the quadratic decay, while for $c>1$, it is shorter.\\
\noindent\textit{Short-time fractional power-law decay with a finite cutoff.—}
Let us now consider the model with a finite cutoff in the dispersion of the environment. The purpose of this cutoff is to examine whether the fractional power-law decay obtained for the unbounded dispersion survives in the presence of a finite cutoff, as a minimal regularization relevant to engineered reservoirs.
 In this section, we focus on the time scale in which terms up to the second order in $g$ are dominant. Keeping terms up to the second order in the Dyson series expansion (\ref{23}), we have
\begin{align}
    \braket{\mS|\psi(t)}\simeq  \me^{-\mi \omega_{\mS}t}&\Bigg[1- g^2\int_{0}^{t} \dd{t_1} \int_{0}^{t_1} \dd{t_2} \me^{\mi \omega_{\mS}(t_1-t_2)}\notag\\
    &~~\int_{|\vec{k}|\leq \Lambda}\dd{\vec{k}} \me^{-\mi \omega_0|\vec{k}|^n (t_1-t_2)}\Bigg]\label{28},
\end{align} 
where $\Lambda$ is the cutoff.
The second-order term of $g$ can be further expanded as follows:
\begin{align}
  S_D\Lambda^D
  \sum_{p=0}^{\infty}\sum_{\ell=0}^{p}
  \frac{(-\omega_0\Lambda^n)^\ell \mi^{p}\omega_{\rm S}^{p-\ell}}{\ell ! (p-\ell)! n(\nu+\ell)(p+1)(p+2)}t^{p+2};\label{finite}
\end{align}
see Sec.~B.1 of Supplemental Material~\cite{supplemental_material}. We note that $S_Df(t)$ therein corresponds to Eq.~(\ref{finite}). In the short-time regime, the most dominant term is given by $p=\ell=0$, which yields the quadratic decay $1-g^2S_D\Lambda^D t^2/(2D)$; see Sec.~B.2 of Supplemental Material~\cite{supplemental_material}. On the other hand, by taking into account all terms with $p=\ell\geq 0$, we can deform Eq.~(\ref{finite}) to 
\begin{gather}
S_D\omega_0^{-\nu}t^{2-\nu}\Bigg[-\dfrac{\mi^{-\nu}\Gamma(\nu-1)}{n(2-\nu)}
+\mathcal{O}\qty((\omega_0\Lambda^n t)^{\nu-1})\Bigg];
\end{gather}
see Sec.~B.3 of Supplemental Material~\cite{supplemental_material}. Therefore, the fractional power-law decay $t^{2-\nu}$ emerges in the time regime $(\omega_0\Lambda^n)^{-1}\ll t<\tau_Z$ with $\nu<1$. The upper bound $\tau_{Z}$ is given because of the same logic as the case without the cutoff.

This fact implies that the fractional power-law dynamics (\ref{30}) in the short-time regime for the infinite cutoff $\Lambda=\infty$ survives in the time scale after the quadratic decay when the cutoff is finite, provided the large cutoff $\omega_0\Lambda^n/\omega_{\mS}\gg 1$, from which the condition $p=\ell$ follows. The large cutoff means that the time scale of the environment $1/(\omega_0\Lambda^n)$ is much shorter than that of the two-level system $1/\omega_{\mS}$.


\noindent\textit{Dynamics in the long-time limit.—}
The Dyson-series expansion is not convenient for analyzing the long-time behavior because it is an expansion in both the coupling strength and time. To extract the long-time asymptotics, we use the Feshbach formalism \cite{feshbach1958unified,hatano2014time}. This formalism rewrites the survival amplitude as a contour integral of the resolvent in the complex-energy plane, in which the pole and branch-cut contributions become explicit. In particular, it separates the exponential-decay contribution, the long-time power-law contribution, and the bound-state contribution. In our model, the survival amplitude can be expressed in terms of the effective Hamiltonian as follows:  
\begin{align}
  &\braket{\mathrm{S}|\psi(t)}=\frac{1}{2\pi \mi} \int_{\infty+\mi \eta}^{-\infty+\mi \eta}\dd{z} \frac{\me^{-\mi z t}}{z -\bra{\mS}H_{\mathrm{eff}}(z)\ket{\mS}},\label{33}\\
  &H_{\mathrm{eff}}(z):= \mP H\mP +\mP H\mQ\frac{1}{z-\mQ H\mQ}\mQ H\mP, \label{21}
\end{align}
where 
\begin{gather}
  \mP:=\ketbra{\mS},~\mQ:= I-\mP=\int_{\mathbb{R}^D}\dd \vec{k} \ketbra*{\vec{k}}  
\end{gather}
 and hence
 \begin{gather}
    \mP H\mP=H_{\mathrm{S}}=\omega_{\mS}\ketbra{\mS},\\
    \mQ H\mQ=H_{\mathrm{E}}=\omega_0\int_{\mathbb{R}^D}\dd \vec{k} |\vec{k}|^n \ketbra*{\vec{k}},\\
\mP H\mQ=g\int_{\mathbb{R}^D}\dd \vec{k}\ketbra*{\mS}{\vec{k}}, \mQ H\mP=g\int_{\mathbb{R}^D}\dd \vec{k}\ketbra*{\vec{k}}{\mS};
 \end{gather}
  see Sec.~C.1 of Supplemental Material~\cite{supplemental_material}. Here, we added an infinitesimal positive imaginary part $\mi \eta$ to the integral contour in Eq.~(\ref{33}) to choose the retarded Green's function for $t>0$.
Then the effective Hamiltonian can be calculated as
\begin{align}
  H_{\mathrm{eff}}(z)
    &=\qty[\omega_{\mS}-g^2 \frac{S_D\omega_0^{-\nu}}{n} \frac{\pi \me^{\mi (1-\nu )\pi }}{\sin[(1-\nu )\pi]} z^{\nu-1}]\ketbra{\mS}\label{36}
\end{align}
for $0<\nu<1$; see Sec.~C.2 of Supplemental Material~\cite{supplemental_material}.
Combining Eqs.~(\ref{33}) and (\ref{36}), we arrive at the survival amplitude in the form
  \begin{gather}
   \braket{\mS|\psi(t)}=\frac{1}{2\pi \mi} \int_{\infty+\mi \eta}^{-\infty+\mi \eta}\dd{z} F(z) \label{37},\\
    F(z):=\frac{\me^{-\mi z t}}
    {z -\omega_{\mS}+B \me^{\mi (1-\nu )\pi } z^{\nu-1}},\\
    B:=g^2 \frac{S_D\omega_0^{-\nu}}{n} \frac{\pi}{\sin[(1-\nu )\pi]},
  \end{gather}  
with $\eta>0$ for $t>0$ and $B>0$ for $0<\nu <1$. Since the integrand involves the multi-valued function $z^{\nu-1}$ with $0<\nu<1$, we analyze its Riemann surfaces. We parametrize the complex variable as $z = r \me^{\mi\theta}~~(r\ge0,\;\theta\in\mathbb{R}),$ choose the branch cut along $[0,-\mi \infty]$, define the first Riemann sheet as $-\pi/2\leq \theta <3\pi/2$, and evaluate the integral in Eq.~(\ref{37}) using the Cauchy residue theorem. 
We take the contour as depicted in Fig. 2.  
\begin{figure}[t]
  \centering
  \includegraphics[width=60mm]{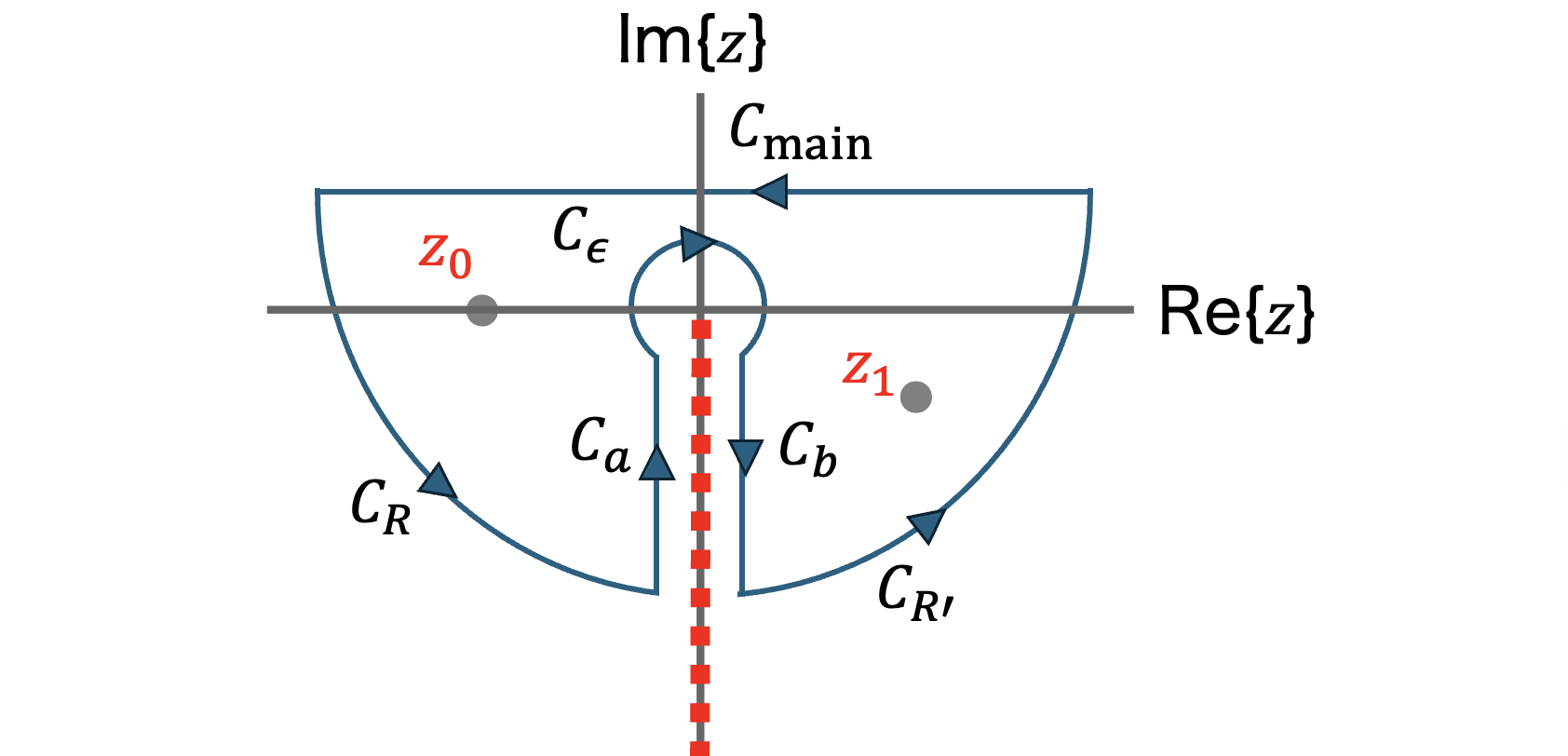}
  \caption{Contour used for the integral in evaluating Eq.~(\ref{37}). The path $C_{\mathrm{main}}$ extending from $\infty +\mi \eta$ to $-\infty+\mi \eta$ represents the integral that we aim to evaluate. We close the integral contour by adding the semicircular arcs $C_{R}$ and $C_{R'}$ with radius $R$, straight paths $C_{a}$ and $C_{b}$ that extend from $\mp \mi \infty \mp\epsilon$ to $0 \mp \epsilon$, respectively, and a small circular $C_{\epsilon}$ with radius $\epsilon$. The point $z_0$ represents the simple pole on the negative real axis of the first Riemann sheet. Depending on the parameters of the model, one simple pole may exist on the fourth quadrant of the first Riemann sheet, which is denoted by $z_1$.} 
\end{figure}
For $-\pi/2 \leq \theta\leq 0$ and $\pi\leq \theta <3\pi/2$, we have $F(R\me^{\mi \theta})\to 0$ in the limit $R\to \infty$ according to Jordan's lemma and $F(\epsilon \me^{\mi \theta})\to 0$ in the limit $\epsilon \to 0$.
 Therefore, the integrals along the path $C_R$, $C_{R'}$ and $C_{\epsilon}$ vanish.
In the first Riemann sheet, there exists a simple pole $z_0$ on the negative real axis $\theta=\pi$ and may or may not exist a simple pole $z_1$ on the fourth quadrant $-\pi/2\le \theta<0$; see Secs.~C.3 and C.4 of Supplemental Material~\cite{supplemental_material}.
\begin{figure}[t]
    \setcounter{panel}{0} 
  \centering
  \begin{minipage}{0.48\columnwidth}
  \includegraphics[width=\linewidth]{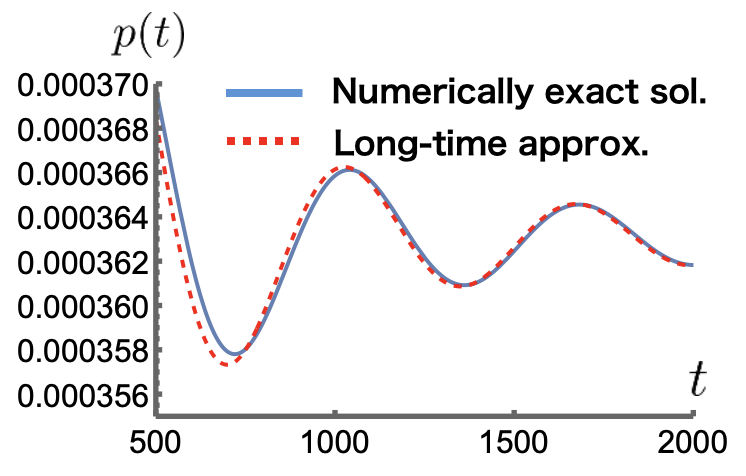}
  \par\small\centering \paneltag{fig:pt-a}
  \end{minipage}\hfill
  \begin{minipage}{0.48\columnwidth}
  \includegraphics[width=\linewidth]{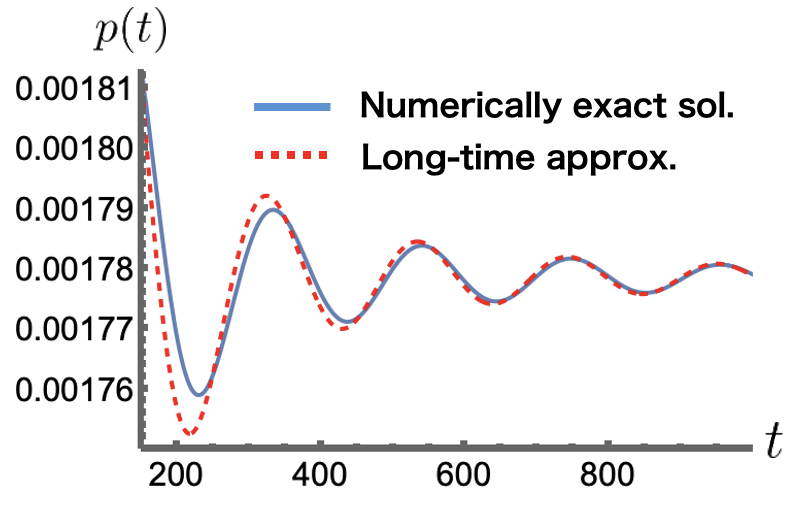}
  \par\small\centering \paneltag{fig:pt-b}
  \end{minipage}
  \caption{Time dependence of the survival probability $p(t)$; numerically exact evaluation of Eq.~({\ref{33}}) (blue solid line) and asymptotic behavior according to Eq.~(\ref{long-survive}) (red dotted line): (a) $\omega_{\mathrm{S}}=1$, $B=0.1$ and $\nu=1/2$; (b) $\omega_{\mathrm{S}}=1$, $B=0.1$ and $\nu=1/3$.}
\end{figure}

The integral in Eq.~(\ref{37}) can therefore be decomposed into the contributions from the branch cut and the residues at the poles:
\begin{align}
  \int _{C_\mathrm{main}}\dd{z} F(z)&= -\int _{C_\mathrm{a}+C_\mathrm{b}}\dd{z} F(z)\notag\\
  &~~~~+2\pi \mi(\Res[F,z_0]+\Res[F,z_1]).\label{49}
\end{align}
where
\begin{gather}
    \Res[F,z_j]= \alpha_j \me^{-\mi z_j t}
\end{gather}
with 
\begin{gather}
    \alpha_j=\qty[1+B(\nu-1)\me^{\mi(1-\nu)\pi}z_j^{\nu-2}]^{-1}
\end{gather}
for $j=0,~1$.
Note, however, that the last term of Eq.~(\ref{49}) for $z_1$ may not exist in some parameter regions.
Since $\Im z_1<0$, the residue of $z_1$ decays exponentially and does not contribute to the long-time dynamics anyway. The residue of $z_0$ survives because it is real.
We calculate the integral along the paths $C_\mathrm{a}+C_\mathrm{b}$ using the asymptotic expansion \cite{garmon2013amplification,taira2024markovianity} in terms of the leading order of $1/t$:
  \begin{align}
    &\int _{C_\mathrm{a}+C_\mathrm{b}}\dd{z} F(z)\notag\\
   &\simeq\frac{\mi t^{\nu-2}}{B}\qty[ \me^{\mi(1-\nu)\pi/2}-\me^{-3\mi(1-\nu)\pi/2} ]\Gamma(2-\nu);\label{57}
  \end{align}
  see Sec.~C.5 of Supplemental Material~\cite{supplemental_material}.
    Collecting the contributions from the residue of $z_0$ and the asymptotic form of the integration along $C_a +C_b$ in Eq.~(\ref{57}), we arrive at the long-time behavior of the survival amplitude in the form
      \begin{align}
        \braket{\mS|\psi(t)}\simeq &c_{\mathrm{S}}(0)\Bigg\{-\frac{ \Gamma(2-\nu)}{2\pi B}\Big[ \me^{\mi(1-\nu)\pi/2}\notag\\
        &-\me^{-3\mi(1-\nu)\pi/2} \Big]t^{\nu-2}+\alpha_0 \me^{-\mi z_0 t}\Bigg\}.\label{long-time}
      \end{align}
    The branch-cut contribution (\ref{57}) exhibits a power-law decay, while the real pole $z_0$ gives it an oscillation. The survival probability is given as follows:
    \begin{align}
        p(t)\simeq &|\alpha_0|^2-\frac{\Gamma(2-\nu)}{\pi B}\mathrm{Re}\Big[\alpha_0\Big(\exp{-\mi\qty[z_0t+(1-\nu)\pi/2]}\notag\\
        &-\exp{-\mi\qty[z_0 t-3(1-\nu)\pi/2]}\Big)\Big]t^{\nu-2}.\label{long-survive}
    \end{align}
    
    Therefore, the long-time behavior of the survival probability deviates from the exponential decay and scales as $t^{\nu-2}$ with oscillation; see Fig.~3 for comparison of the asymptotic behavior (\ref{long-survive}) with the numerically exact calculation of Eq.~(\ref{33}). 

\noindent\textit{Conclusion and discussion.—}In this paper, we analyzed the dynamics of the two-level system coupled with the environment with dispersion $\omega(k)=\omega_0|k|^n$ in $D$ spatial dimensions. In the short-time regime, we employed the Dyson series expansion to obtain the survival amplitude $\braket{\mS|\psi(t)}$ and found that it scales as $1- ct^{2-D/n}$ in $0<D/n<1$ for small $t$, where $c$ is a constant. Hence, the survival probability in the short-time regime induces the anomalous quantum Zeno effect. Notably, other works have predicted the possibility of a short-time decay of the form $1-ct^{3/2}$, provided that the quantity $\braket{H^2}-\braket{H}^2$ diverges \cite{cordero2012analytical,muga1996survival}. 

The origin of the short-time decay with a fractional exponent is the two-point autocorrelation function of the environment, which plays a role of the memory kernel. Our perspective from the vantage point of the memory kernel is different from approaches characterizing non-Markovianity at the level of the reduced dynamical map, such as CP-divisibility or information backflow \cite{li2018concepts,shrikant2023quantum}. While these approaches characterize properties of the reduced dynamics, the memory kernel directly captures how past states influence the present dynamics, providing a physical picture of the breakdown of the semigroup description.

If the energy spectrum has an upper bound $\Lambda$, the very short-time behavior of the survival probability is quadratic, but it crosses over to the fractional power-law decay around $t\sim (\omega_0\Lambda^n)^{-1}$.
 This suggests that the fractional power-law decay may be accessible in an engineered environment with a sufficiently large but a finite upper cutoff.

We also examined the long-time dynamics using the Feshbach effective Hamiltonian (\ref{36}) and found that the survival probability scales as $c t^{\nu-2}$ for large $t$ with oscillations as shown in Eq.~(\ref{long-time}).

Finally, we clarify the relation between the present result and the regime $\nu\geq 1$.
The present analysis is restricted to $0<\nu<1$. This restriction is not merely technical; in the unregularized model with lower-bounded but unbounded-above dispersion and momentum-independent coupling, both the environmental autocorrelation function~(\ref{11}) and the Feshbach effective Hamiltonian~(\ref{21}) are UV divergent for $\nu\geq 1$. Thus, our formulas cannot be directly extended to this regime. Note that this regime can be analyzed if a finite cutoff is introduced. Indeed, previous studies of photonic reservoirs, including the three-dimensional linear-dispersion case, have treated such situations using a cutoff, a momentum-dependent coupling constant, or a finite bandwidth, where non-Markovian radiation dynamics and deviations from simple exponential decay can occur \cite{seke1988deviations,berman2010spontaneous,grimsmo2013memory}.



\begin{acknowledgments}
  H.N. was supported by the WINGS-QSTEP Program, the University of Tokyo. H.K. was supported by JST SPRING, Grant No.~JPMJSP2108. T.T. was supported by JSPS Grant No.~22J01230 and R7 (2025) Young Researchers Support Project, Faculty of Science, KYUSHU UNIVERSITY. N.H. was supported by JSPS Grant No.~JP24K00545 and JP23K22411.
  \end{acknowledgments}
  \bibliographystyle{apsrev4-2}
  \nocite{*}
  \bibliography{main_prl}

\clearpage
\onecolumngrid

\begin{center}
  {\large\bfseries Supplemental Material for ``Fractional power-law decay in the spontaneous emission of a two-level system''\par}
  \vspace{1em}
  Hiroki Nakabayashi,$^{1,*}$ Hayato Kinkawa,$^{1,\dagger}$ Takano Taira,$^{2,3,\ddagger}$ and Naomichi Hatano$^{3,\mathsection}$

  \vspace{0.5em}
  {\itshape
  $^1$Department of Physics, The University of Tokyo, 5-1-5 Kashiwanoha, Kashiwa, Chiba 277-8574, Japan\\
  $^2$Department of Physics, Kyushu University, 744 Motooka, Nishiku, Fukuoka 819-0395, Japan\\
  $^3$Institute of Industrial Science, The University of Tokyo, 5-1-5 Kashiwanoha, Kashiwa, Chiba 277-8574, Japan}
\end{center}

\begingroup
  \setcounter{secnumdepth}{0}

  \let\mainlabel\label
  \let\mainref\ref
  \renewcommand{\label}[1]{\mainlabel{SM:#1}}
  \renewcommand{\ref}[1]{\mainref{SM:#1}}
  \makeatletter
  \let\mainlabelindisplay\label@in@display
  \def\label@in@display#1{\mainlabelindisplay{SM:#1}}

  \def\SMbibcite#1#2{\expandafter\def\csname b@#1\endcsname{#2}}
  \SMbibcite{gradshteyn2014table}{{1}{2014}{{Gradshteyn\ and\ Ryzhik}}{{}}}
  \SMbibcite{facchi2008quantum}{{2}{2008}{{Facchi\ and\ Pascazio}}{{}}}
  \SMbibcite{feshbach1958unified}{{3}{1958}{{Feshbach}}{{}}}
  \SMbibcite{hatano2014time}{{4}{2014}{{Hatano\ and\ Ordonez}}{{}}}
  \SMbibcite{garmon2013amplification}{{5}{2013}{{Garmon\ \emph{et~al.}}}{{Garmon, Petrosky, Simine,\ and\ Segal}}}
  \SMbibcite{taira2024markovianity}{{6}{2024}{{Taira\ \emph{et~al.}}}{{Taira, Hatano,\ and\ Nishino}}}
  \SMbibcite{breuer2002theory}{{7}{2002}{{Breuer\ and\ Petruccione}}{{}}}
  \SMbibcite{garraway1997nonperturbative}{{8}{1997}{{Garraway}}{{}}}
  \makeatother

  \def\preprint#1{}
  \def\title#1{}
  \def\author#1{}
  \def\email#1{}
  \def\affiliation#1{}
  \def\maketitle{}
  \def\nocite#1{}
  \def\bibliographystyle#1{}
  \makeatletter
  \def\bibliography#1{%
    \begingroup
      \@fileswfalse
\begin{thebibliography}{8}%
\makeatletter
\providecommand \@ifxundefined [1]{%
 \@ifx{#1\undefined}
}%
\providecommand \@ifnum [1]{%
 \ifnum #1\expandafter \@firstoftwo
 \else \expandafter \@secondoftwo
 \fi
}%
\providecommand \@ifx [1]{%
 \ifx #1\expandafter \@firstoftwo
 \else \expandafter \@secondoftwo
 \fi
}%
\providecommand \natexlab [1]{#1}%
\providecommand \enquote  [1]{``#1''}%
\providecommand \bibnamefont  [1]{#1}%
\providecommand \bibfnamefont [1]{#1}%
\providecommand \citenamefont [1]{#1}%
\providecommand \href@noop [0]{\@secondoftwo}%
\providecommand \href [0]{\begingroup \@sanitize@url \@href}%
\providecommand \@href[1]{\@@startlink{#1}\@@href}%
\providecommand \@@href[1]{\endgroup#1\@@endlink}%
\providecommand \@sanitize@url [0]{\catcode `\\12\catcode `\$12\catcode
  `\&12\catcode `\#12\catcode `\^12\catcode `\_12\catcode `\%12\relax}%
\providecommand \@@startlink[1]{}%
\providecommand \@@endlink[0]{}%
\providecommand \url  [0]{\begingroup\@sanitize@url \@url }%
\providecommand \@url [1]{\endgroup\@href {#1}{\urlprefix }}%
\providecommand \urlprefix  [0]{URL }%
\providecommand \Eprint [0]{\href }%
\providecommand \doibase [0]{https://doi.org/}%
\providecommand \selectlanguage [0]{\@gobble}%
\providecommand \bibinfo  [0]{\@secondoftwo}%
\providecommand \bibfield  [0]{\@secondoftwo}%
\providecommand \translation [1]{[#1]}%
\providecommand \BibitemOpen [0]{}%
\providecommand \bibitemStop [0]{}%
\providecommand \bibitemNoStop [0]{.\EOS\space}%
\providecommand \EOS [0]{\spacefactor3000\relax}%
\providecommand \BibitemShut  [1]{\csname bibitem#1\endcsname}%
\let\auto@bib@innerbib\@empty
\bibitem [{\citenamefont {Gradshteyn}\ and\ \citenamefont
  {Ryzhik}(2014)}]{gradshteyn2014table}%
  \BibitemOpen
  \bibfield  {author} {\bibinfo {author} {\bibfnamefont {I.~S.}\ \bibnamefont
  {Gradshteyn}}\ and\ \bibinfo {author} {\bibfnamefont {I.~M.}\ \bibnamefont
  {Ryzhik}},\ }\href@noop {} {\emph {\bibinfo {title} {{Table of integrals,
  series, and products}}}}\ (\bibinfo  {publisher} {Academic press},\ \bibinfo
  {year} {2014})\BibitemShut {NoStop}%
\bibitem [{\citenamefont {Facchi}\ and\ \citenamefont
  {Pascazio}(2008)}]{facchi2008quantum}%
  \BibitemOpen
  \bibfield  {author} {\bibinfo {author} {\bibfnamefont {P.}~\bibnamefont
  {Facchi}}\ and\ \bibinfo {author} {\bibfnamefont {S.}~\bibnamefont
  {Pascazio}},\ }\bibfield  {title} {\bibinfo {title} {{Quantum Zeno dynamics:
  mathematical and physical aspects}},\ }\href@noop {} {\bibfield  {journal}
  {\bibinfo  {journal} {Journal of Physics A: Mathematical and Theoretical}\
  }\textbf {\bibinfo {volume} {41}},\ \bibinfo {pages} {493001} (\bibinfo
  {year} {2008})}\BibitemShut {NoStop}%
\bibitem [{\citenamefont {Feshbach}(1958)}]{feshbach1958unified}%
  \BibitemOpen
  \bibfield  {author} {\bibinfo {author} {\bibfnamefont {H.}~\bibnamefont
  {Feshbach}},\ }\bibfield  {title} {\bibinfo {title} {{Unified theory of
  nuclear reactions}},\ }\href@noop {} {\bibfield  {journal} {\bibinfo
  {journal} {Annals of Physics}\ }\textbf {\bibinfo {volume} {5}},\ \bibinfo
  {pages} {357} (\bibinfo {year} {1958})}\BibitemShut {NoStop}%
\bibitem [{\citenamefont {Hatano}\ and\ \citenamefont
  {Ordonez}(2014)}]{hatano2014time}%
  \BibitemOpen
  \bibfield  {author} {\bibinfo {author} {\bibfnamefont {N.}~\bibnamefont
  {Hatano}}\ and\ \bibinfo {author} {\bibfnamefont {G.}~\bibnamefont
  {Ordonez}},\ }\bibfield  {title} {\bibinfo {title} {{Time-reversal symmetric
  resolution of unity without background integrals in open quantum systems}},\
  }\href@noop {} {\bibfield  {journal} {\bibinfo  {journal} {Journal of
  Mathematical Physics}\ }\textbf {\bibinfo {volume} {55}} (\bibinfo {year}
  {2014})}\BibitemShut {NoStop}%
\bibitem [{\citenamefont {Garmon}\ \emph {et~al.}(2013)\citenamefont {Garmon},
  \citenamefont {Petrosky}, \citenamefont {Simine},\ and\ \citenamefont
  {Segal}}]{garmon2013amplification}%
  \BibitemOpen
  \bibfield  {author} {\bibinfo {author} {\bibfnamefont {S.}~\bibnamefont
  {Garmon}}, \bibinfo {author} {\bibfnamefont {T.}~\bibnamefont {Petrosky}},
  \bibinfo {author} {\bibfnamefont {L.}~\bibnamefont {Simine}},\ and\ \bibinfo
  {author} {\bibfnamefont {D.}~\bibnamefont {Segal}},\ }\bibfield  {title}
  {\bibinfo {title} {{Amplification of non-Markovian decay due to bound state
  absorption into continuum}},\ }\href@noop {} {\bibfield  {journal} {\bibinfo
  {journal} {Fortschritte der Physik}\ }\textbf {\bibinfo {volume} {61}},\
  \bibinfo {pages} {261} (\bibinfo {year} {2013})}\BibitemShut {NoStop}%
\bibitem [{\citenamefont {Taira}\ \emph {et~al.}(2024)\citenamefont {Taira},
  \citenamefont {Hatano},\ and\ \citenamefont
  {Nishino}}]{taira2024markovianity}%
  \BibitemOpen
  \bibfield  {author} {\bibinfo {author} {\bibfnamefont {T.}~\bibnamefont
  {Taira}}, \bibinfo {author} {\bibfnamefont {N.}~\bibnamefont {Hatano}},\ and\
  \bibinfo {author} {\bibfnamefont {A.}~\bibnamefont {Nishino}},\ }\bibfield
  {title} {\bibinfo {title} {{Markovianity and non-Markovianity of Particle
  Bath with Dirac Dispersion Relation}},\ }\href@noop {} {\bibfield  {journal}
  {\bibinfo  {journal} {arXiv preprint arXiv:2406.17436}\ } (\bibinfo {year}
  {2024})}\BibitemShut {NoStop}%
\bibitem [{\citenamefont {Breuer}\ and\ \citenamefont
  {Petruccione}(2002)}]{breuer2002theory}%
  \BibitemOpen
  \bibfield  {author} {\bibinfo {author} {\bibfnamefont {H.-P.}\ \bibnamefont
  {Breuer}}\ and\ \bibinfo {author} {\bibfnamefont {F.}~\bibnamefont
  {Petruccione}},\ }\href@noop {} {\emph {\bibinfo {title} {{The theory of open
  quantum systems}}}}\ (\bibinfo  {publisher} {OUP Oxford},\ \bibinfo {year}
  {2002})\BibitemShut {NoStop}%
\bibitem [{\citenamefont {Garraway}(1997)}]{garraway1997nonperturbative}%
  \BibitemOpen
  \bibfield  {author} {\bibinfo {author} {\bibfnamefont {B.~M.}\ \bibnamefont
  {Garraway}},\ }\bibfield  {title} {\bibinfo {title} {{Nonperturbative decay
  of an atomic system in a cavity}},\ }\href@noop {} {\bibfield  {journal}
  {\bibinfo  {journal} {Physical Review A}\ }\textbf {\bibinfo {volume} {55}},\
  \bibinfo {pages} {2290} (\bibinfo {year} {1997})}\BibitemShut {NoStop}%
\end{thebibliography}%
    \endgroup
  }
  \makeatother

\preprint{APS/123-QED}

\title{Supplemental Material for “Fractional power-law decay in the spontaneous emission of a two-level system”}

\author{Hiroki Nakabayashi}
\email{nakaba@iis.u-tokyo.ac.jp}
\affiliation{Department of Physics, The University of Tokyo, 5-1-5 Kashiwanoha, Kashiwa, Chiba 277-8574, Japan}
\author{Hayato Kinkawa}
\email{kinkawa@iis.u-tokyo.ac.jp}
\affiliation{Department of Physics, The University of Tokyo, 5-1-5 Kashiwanoha, Kashiwa, Chiba 277-8574, Japan}
\author{Takano Taira}
\email{taira.takano.292@m.kyushu-u.ac.jp}
\affiliation{Department of Physics, Kyushu University, 744 Motooka, Nishiku, Fukuoka 819-0395, Japan}
\author{Naomichi Hatano}
\email{hatano@iis.u-tokyo.ac.jp}
\affiliation{Institute of Industrial Science, The University of Tokyo, 5-1-5 Kashiwanoha, Kashiwa, Chiba 277-8574, Japan}


\maketitle
\setcounter{equation}{0}
\renewcommand{\theequation}{S\arabic{equation}}

\setcounter{figure}{0}
\renewcommand{\thefigure}{S\arabic{figure}}

\setcounter{table}{0}
\renewcommand{\thetable}{S\arabic{table}}

\section{A.~Derivation of Eq.~(13)}
We here derive a series expansion of the survival amplitude of the excited state of Eq.~(13) in the main text. We recall that the Hamiltonian of the total system in the first quantization form is given by Eq.~(8) as follows:
\begin{gather}
    H =H_0+H_{\mathrm{int}},\label{S1}\\
     H_0=\omega_{\mathrm{S}} \ketbra{\mathrm{S}}{\mathrm{S}} +\omega_0 \int_{\mathbb{R}^D}\dd{\vec{k}} |\vec{k}|^{n} \ketbra*{\vec{k}},\\
     H_{\mathrm{int}}=g \int_{\mathbb{R}^D}\dd{\vec{k}} (\ketbra*{\mathrm{S}}{\vec{k}}  + \ketbra*{\vec{k}}{\mathrm{S}} ),\label{15}
  \end{gather}
and the initial state of the time evolution is the excited state $\ket{\mS}$.
\subsection{A.1 Dyson series expansion}
In order to obtain the series solution of the survival amplitude of the excited state, we use the Dyson series expansion for $\ket{\psi(t)}$ of Eq.~(12) in the main text as follows:
\begin{align}
    &\braket{\mS|\psi(t)}=\me^{-\mi \omega_{\mS}t}\sum_{\ell =0}^{\infty} \qty(-\mi)^{\ell} \int_{0}^{t} \dd{t_1} \int_{0}^{t_1} \dd{t_2} \cdots \int_{0}^{t_{\ell -1}} \dd{t_{\ell}} \braket{\mS|H_{\mathrm{int}}^{(\rm I)}(t_1)H_{\mathrm{int}}^{(\rm I)}(t_2)\cdots H_{\mathrm{int}}^{(\rm I)}(t_{\ell})|\mS},\label{S2}
\end{align}
with the coupling Hamiltonian in the interaction picture
\begin{align}
    H_{\mathrm{int}}^{(\rm I)}(t) &=\me^{\mi H_0 t}H_{\mathrm{int}}\me^{-\mi H_0 t} \\
    &=\qty(\me^{\mi \omega_{\mS}t}\ketbra*{\mS}+\int_{\mathbb{R}^D}\dd{\vec{k}} \me^{\mi \omega_0|\vec{k}|^n t}\ketbra*{\vec{k}})\notag\\
    &~~~~g\int_{\mathbb{R}^D}\dd{\vec{k}}\qty(\ketbra*{\mS}{\vec{k}}+\ketbra*{\vec{k}}{\mS})\notag\\
    &~~~~\qty(\me^{-\mi \omega_{\mS}t}\ketbra*{\mS}+\int_{\mathbb{R}^D}\dd{\vec{k}} \me^{-\mi \omega_0|\vec{k}|^n t}\ketbra*{\vec{k}})\\
    &=g\qty{\int_{\mathbb{R}^D}\dd{\vec{k}} [\me^{\mi (\omega_{\mathrm{S}}-\omega_0|\vec{k}|^n ) t}\ketbra*{\mS}{\vec{k}} + \mathrm{h.c.} ]}.
\end{align}
For brevity, we absorb the phase factor $\me^{\mi \omega_{\mS}t}$ and define the survival amplitude in the interaction picture as follows:
\begin{gather}
    \braket{\mS|\psi^{\mathrm{I}}(t)}:=\me^{\mi \omega_{\mS}t}\braket{\mS|\psi(t)}.\label{S7}
\end{gather}

First, let us calculate the integrand of Eq.~(\ref{S2}). Only the terms in which the interaction Hamiltonian is multiplied even times survive due to the off-diagonal structure of $H_{\mathrm{int}}^{(\rm I)}(t)$ in the $\{\ket{\mS},\ket{\vec k}\}$ basis. In particular, we have
\begin{align}
H_{\mathrm{int}}^{(\rm I)}(t_1)H_{\mathrm{int}}^{(\rm I)}(t_2)&=g^2\int_{\mathbb{R}^D}\dd{\vec{k}} \me^{\mi (\omega_{\mS}-\omega_0|\vec{k}|^n ) (t_1-t_2)}\ketbra{\mS}\notag\\
&~~~~+g^2\int_{\mathbb{R}^D}\dd{\vec{k}} \int_{\mathbb{R}^D} \dd{\vec{k}'} \me^{\mi (\omega_{\mS}-\omega_0|\vec{k}|^n ) t_2}\me^{-\mi (\omega_{\mS}-\omega_0|\vec{k'}|^n  ) t_1}\ketbra*{\vec{k}'}{\vec{k}}.\label{S9}
\end{align}
Then the integrand in Eq.~(\ref{S2}) can be rewritten as follows:
\begin{align}
    \braket{\mS|H_{\mathrm{int}}^{(\rm I)}(t_1)H_{\mathrm{int}}^{(\rm I)}(t_2)\cdots H_{\mathrm{int}}^{(\rm I)}(t_{2\ell})|\mS}
    &=\braket{\mS|H_{\mathrm{int}}^{(\rm I)}(t_1)H_{\mathrm{int}}^{(\rm I)}(t_2)|\mS}\cdots\braket{\mS|H_{\mathrm{int}}^{(\rm I)}(t_{2\ell-1})H_{\mathrm{int}}^{(\rm I)}(t_{2\ell})|\mS}\\
    &=
    \prod_{j=1}^{\ell} \eta(t_{2j-1}-t_{2j}),\label{S5}\\
    \braket{\mS|H_{\mathrm{int}}^{(\rm I)}(t_1)H_{\mathrm{int}}^{(\rm I)}(t_2)\cdots H_{\mathrm{int}}^{(\rm I)}(t_{2\ell+1})|\mS} &=0,\label{S6}
\end{align}
where we have defined the autocorrelation $\eta(t_1-t_2)$ of the form
\begin{align}
    \eta(t_1-t_2):=\braket{\mS|H_{\mathrm{int}}^{(\rm I)}(t_1)H_{\mathrm{int}}^{(\rm I)}(t_2)|\mS},\label{S12!}
\end{align}
which is the same as Eq.~(15) in the main text.
Therefore, the survival amplitude (\ref{S7}) is given by 
\begin{align}
    \braket{\mS|\psi^{\mathrm{I}}(t)}&=\sum_{\ell =0}^{\infty} \qty(-\mi)^{\ell} \int_{0}^{t} \dd{t_1} \int_{0}^{t_1} \dd{t_2} \cdots \int_{0}^{t_{\ell -1}} \dd{t_{\ell}} \braket{\mS|H_{\mathrm{int}}^{(\rm I)}(t_1)H_{\mathrm{int}}^{(\rm I)}(t_2)\cdots H_{\mathrm{int}}^{(\rm I)}(t_{\ell})|\mS}\\
    &=\sum_{\ell =0}^{\infty} \qty(-\mi)^{2\ell} \int_{0}^{t} \dd{t_1} \int_{0}^{t_1} \dd{t_2} \cdots \int_{0}^{t_{2\ell -1}} \dd{t_{2\ell}}\prod_{j=1}^{\ell} \eta(t_{2j-1}-t_{2j}). \label{dyson}
\end{align}
\subsection{A.2 Calculation of the autocorrelation function}
Next, we show that the autocorrelation function (\ref{S12!}) is given by
\begin{gather}
    \eta(t_1-t_2)=g^2\me^{\mi \omega_{\mS}(t_1-t_2)}\frac{S_D \omega_0^{-\nu} }{n}\me^{-\mi\nu\pi/2}\Gamma(\nu)\qty(t_1-t_2)^{-\nu},\label{S15!}
\end{gather}
where $S_D$ denotes the $D-1$ dimensional surface area of the $D$-dimensional unit sphere, $\nu=D/n$, and $\Gamma(\cdot)$ is the gamma function. We note $\eta(t_1-t_2)\propto (t_1-t_2)^{-\nu}$ as stressed below Eq.~(15) in the main text.
 We present the calculation below. 
 
 Since only the first $\ketbra*{\mS}$ term of Eq.~(\ref{S9}) contributes to the autocorrelation function (\ref{S12!}), we have 
 \begin{gather}
    \eta(t_1-t_2)=g^2\int_{\mathbb{R}^D}\dd{\vec{k}} \me^{\mi (\omega_{\mS}-\omega_0|\vec{k}|^n ) (t_1-t_2)}.
 \end{gather}
 We compute it in the $D$-dimensional polar coordinates, taking $k$ as the radial variable as follows:
\begin{align}
   \eta(t_1-t_2)
    &=g^2\me^{\mi \omega_{\mS}(t_1-t_2)}\int_{\mathbb{R}^D}\dd \vec{k} \me^{-\mi \omega_0|\vec{k}|^n  (t_1-t_2)}\\
    &= g^2\me^{\mi \omega_{\mS}(t_1-t_2)}\int_0^\infty \dd{k} S_D k^{D-1} \me^{-\mi \omega_0k^{n}  (t_1-t_2)}.
    \shortintertext{Setting $u=\omega_0k^{n}  (t_1-t_2)$, we find}
    \eta(t_1-t_2)
    &= g^2\me^{\mi \omega_{\mS}(t_1-t_2)}\frac{S_D }{n} \qty[\frac{1}{\omega_0(t_1-t_2)}]^{1/n}\int_0^\infty \dd{u} u^{1/n-1} \qty[\frac{u}{\omega_0(t_1-t_2)}]^{\frac{D-1}{n} } \me^{-\mi u}\\
    &=g^2\me^{\mi \omega_{\mS}(t_1-t_2)}\frac{S_D }{n} \qty[\omega_0(t_1-t_2)]^{-\nu}\int_0^\infty \dd{u}  u^{\nu-1}   \me^{-\mi u},
    \shortintertext{where we set $\nu=D/n$. Then we expand the factor $\me^{-\mi u}$ to trigonometric functions as follows:}
    \eta(t_1-t_2)&=g^2\me^{\mi \omega_{\mS}(t_1-t_2)}\frac{S_D \omega_0^{-\nu}}{n} \qty(t_1-t_2)^{-\nu}\int_0^\infty \dd{u}  u^{\nu-1}   \qty(\cos u- \mi \sin u)\\
    &=g^2\me^{\mi \omega_{\mS}(t_1-t_2)}\frac{S_D \omega_0^{-\nu} }{n} \qty[\Gamma(\nu)\cos(\nu\pi/2)-\mi \Gamma(\nu)\sin(\nu\pi/2)]\qty(t_1-t_2)^{-\nu}\\
    &=g^2\me^{\mi \omega_{\mS}(t_1-t_2)}\frac{S_D \omega_0^{-\nu} }{n}\me^{-\mi\nu\pi/2}\Gamma(\nu)\qty(t_1-t_2)^{-\nu},\label{S15}
\end{align}
where we used the formula of the gamma function for $0<\nu<1$ \cite{gradshteyn2014table}. We thus obtained Eq.~(\ref{S15!}).
Inserting it into Eq.~(\ref{dyson}), we obtain 
\begin{gather}
  \braket{\mathrm{S}|\psi^{\mathrm{I}}(t)}=\sum_{\ell=0}^{\infty} (-\mi)^{2\ell}\int_{0}^{t} \dd{t_1} \int_{0}^{t_1} \dd{t_2} \cdots \int_{0}^{t_{2\ell -1}}\dd{t_{2\ell}}\prod_{j=1}^{\ell} \eta(t_{2j-1}-t_{2j}),\label{S25}
\end{gather}
where
\begin{gather}
  \eta(t_{2j-1}-t_{2j})=g^2\me^{\mi \omega_{\mS}(t_{2j-1}-t_{2j})}\frac{S_D{\omega_0}^{-\nu} }{n} \me^{-\mi\nu\pi/2}\Gamma(\nu)\qty(t_{2j-1}-t_{2j})^{-\nu}.
\end{gather}
\subsection{A.3 Calculation of the higher order Dyson series using the Laplace transform} 
%
In order to calculate the integral (\ref{S25}), we now utilize the Laplace transform. Under the assumption of the possibility of term-by-term Laplace transform, the Laplace transform $ \mathrm{Lap}[\cdot]$ of the survival amplitude (\ref{S25}) reads as follows:
\begin{align}
    \mathrm{Lap}\qty[\braket{\mS|\psi^{\mathrm{I}}(t)}]
    &=\sum_{\ell=0}^{\infty} \qty(-1)^\ell \mathrm{Lap}\qty[h_{\ell}(t)],\label{S26}
\end{align}
where
\begin{gather}
    h_{\ell}(t):=\int_{0}^{t} \dd{t_1} \int_{0}^{t_1} \dd{t_2} \cdots \int_{0}^{t_{2\ell -1}}\dd{t_{2\ell}}\prod_{j=1}^{\ell} \eta(t_{2j-1}-t_{2j}).
\end{gather}

We now prove
\begin{gather}
    \mathrm{Lap}\qty[h_{\ell}(t)]=\frac{1}{s^{\ell+1}}\qty(\mathrm{Lap}[\eta(t)])^\ell,\label{S31}
\end{gather} 
where $s$ is the Laplace transform variable. We first note
\begin{align}
    h_{\ell}(t)
     &=\int_{0}^{t} \dd{t_1} \int_{0}^{t_1} \dd{t_2}\eta(t_{1}-t_{2})\int_{0}^{t_2} \dd{t_3} \int_{0}^{t_3} \dd{t_4}\eta(t_3-t_4)\cdots \int_{0}^{t_{2\ell -2}}\dd{t_{2\ell-1}}\int_{0}^{t_{2\ell -1}}\dd{t_{2\ell}} \eta(t_{2j-1}-t_{2j})\\
    &=\int_{0}^{t} \dd{t_1} \int_{0}^{t_1} \dd{t_2}\eta(t_{1}-t_{2})h_{\ell-1}(t_2).
\end{align}
 Then using the property that the Laplace transform of an integral over $t$ yields an extra factor $1/s$ and the convolution theorem, we have
\begin{align}
    \mathrm{Lap}\qty[h_\ell(t)]
    &=\mathrm{Lap}\qty[\int_{0}^{t} \dd{t_1} \int_{0}^{t_1} \dd{t_2}\eta(t_{1}-t_{2})h_{\ell-1}(t_2)]\\
    &=\frac{1}{s} \mathrm{Lap}\qty[ \int_{0}^{t} \dd{t_2}\eta(t-t_{2})h_{\ell-1}(t_2)]\\
    &=\frac{1}{s} \mathrm{Lap}\qty[\eta(t)]\mathrm{Lap}\qty[h_{\ell-1}(t)].
\end{align}
Therefore, we recursively obtain 
\begin{align}
    \mathrm{Lap}\qty[h_\ell(t)]
     &=\qty(\frac{1}{s} \mathrm{Lap}\qty[\eta(t)])^{\ell-1}\mathrm{Lap}\qty[h_{1}(t)].
\end{align}
Since we have
\begin{align}
     \mathrm{Lap}\qty[h_1(t)]&=\mathrm{Lap}\qty[\int_{0}^{t} \dd{t_1} \int_{0}^{t_1} \dd{t_2}\eta(t_{1}-t_{2})]\\
     &=\frac{\mathrm{Lap}\qty[\eta(t)]}{s^2},
\end{align}
we arrive at Eq.~(\ref{S31}).

Next, we calculate the Laplace transform of the autocorrelation function as follows: 
\begin{align}
    \Lap[\eta(t)]&=\Lap\qty[g^2\me^{\mi \omega_{\mS}t} \frac{S_D {\omega_0}^{-\nu}}{n}\me^{-\mi\nu\pi/2}\Gamma(\nu)t^{-\nu}]\\
    &=g^2\frac{S_D{\omega_0}^{-\nu} }{n}\me^{-\mi\nu\pi/2}\Gamma(\nu)\Lap[\me^{\mi \omega_{\mS}t}t^{-\nu}]\\
    &=g^2\frac{S_D{\omega_0}^{-\nu} }{n}\me^{-\mi\nu\pi/2}\Gamma(\nu)\Gamma(1-\nu)(s-\mi \omega_{\mS})^{\nu-1}.\label{Lap}
\end{align}

Therefore, combining Eqs.~(\ref{S26}), (\ref{S31}) and (\ref{Lap}), we obtain the Laplace transform of $\braket{\mS|\psi^{\mathrm{I}}(t)}$ as follows:
\begin{align}
    \mathrm{Lap}[\braket{\mS|\psi^{\mathrm{I}}(t)}]
   &=\sum_{\ell=0}^{\infty} \qty(-g^2)^\ell\frac{1}{s^{\ell+1}}
   \qty[\frac{S_D{\omega_0}^{-\nu} }{n}\me^{-\mi\nu\pi/2}\Gamma(\nu)\Gamma(1-\nu)]^\ell(s-\mi \omega_{\mS})^{\ell \nu-\ell}.
\end{align}
We thus find the survival amplitude using the inverse Laplace transform as follows:
\begin{align}
    \braket{\mS|\psi(t)}&=
    \me^{-\mi \omega_{\mS}t}\braket{\mS|\psi^{\mathrm{I}}(t)}\\
    &=\me^{-\mi \omega_{\mS}t}\sum_{\ell=0}^{\infty} \qty(-g^{2})^\ell
   \qty[ \frac{S_D{\omega_0}^{-\nu} }{n}\me^{-\mi\nu\pi/2}\Gamma(\nu)\Gamma(1-\nu)]^\ell \Lap^{-1}\qty[\frac{1}{s^{\ell+1}}(s-\mi \omega_{\mS})^{\ell \nu-\ell}].\label{S28}
\end{align}
\subsection{A.4 Inverse Laplace transform}
Finally, we derive the inverse Laplace transform in Eq.~(\ref{S28}) and obtain the final form (13) in the main text. 

We apply the convolution theorem for the inverse Laplace transform:
\begin{align}
    \Lap^{-1}\qty[\frac{1}{s^{\ell+1}}(s-\mi \omega_{\mS})^{\ell \nu-\ell}]
    &=\Lap^{-1}\cdot\Lap\qty[\int_0^t \dd{\tau} \frac{(t-\tau)^\ell}{\ell !}\frac{\tau^{\ell-\ell \nu-1}}{\Gamma(\ell -\ell \nu)}\me^{\mi \omega_{\mS}\tau}]\label{inverse}\\
    &=\int_0^t \dd{\tau} \frac{(t-\tau)^\ell}{\ell !}\frac{\tau^{\ell-\ell \nu-1}}{\Gamma(\ell -\ell \nu)}\me^{\mi \omega_{\mS}\tau}. \label{S30}
\end{align}
Introducing the change of variable $\tau= vt$, we obtain
\begin{align}
    (\text{r.h.s. of \ref{S30}})&=\int_0^1 \dd{v} t \frac{(t-vt)^\ell}{\ell!}\frac{(vt)^{\ell-\ell\nu-1}}{\Gamma(\ell-\ell\nu)}\me^{\mi \omega_{\mS}vt}\\
    &=\frac{t^{2\ell-\ell\nu}}{\ell!\Gamma(\ell-\ell\nu)}\int_0^1 \dd{v}  {(1-v)^\ell}{v^{\ell-\ell\nu-1}}\me^{\mi \omega_{\mS}vt}.\label{S32}
\end{align}
The remaining integral can be expressed in terms of the confluent hypergeometric function of the first kind. In fact, using the formula
\begin{gather}
    \int_0^1 \dd{v}  {(1-v)^{b-a-1}}{v^{a-1}}\me^{zv}=\frac{\Gamma(b-a)\Gamma(a)}{\Gamma(b)}{}_1F_1(a,b,z)
\end{gather}
and identifying $a=\ell-\ell \nu$ and $b=2\ell-\ell \nu +1$, we deduce that
\begin{gather}
\int_0^1 \dd{v}  {(1-v)^\ell}{v^{\ell-\ell\nu-1}}\me^{\mi \omega_{\mS}vt}=\frac{\Gamma(\ell-\ell\nu)\Gamma(\ell+1)}{\Gamma(2\ell-\ell\nu+1)}{}_1F_1(\ell-\ell\nu,2\ell-\ell\nu+1,\mi \omega_{\mS}t).\label{S34}
\end{gather}
Therefore the inverse Laplace transform of Eq.~(\ref{inverse}) becomes
\begin{align}
    (\text{l.h.s. of \ref{inverse}})
    &=\frac{t^{2\ell-\ell\nu}}{\ell!\Gamma(\ell-\ell\nu)}\frac{\Gamma(\ell-\ell\nu)\Gamma(\ell+1)}{\Gamma(2\ell-\ell\nu+1)}{}_1F_1(\ell-\ell\nu,2\ell-\ell\nu+1,\mi \omega_{\mS}t)\\
    &=\frac{{}_1F_1(\ell-\ell\nu,2\ell-\ell\nu+1,\mi \omega_{\mS}t)}{\Gamma(2\ell-\ell\nu+1)} t^{2\ell-\ell\nu},\label{S59}
\end{align}
where we have used the formula $\Gamma(\ell+1)=\ell !$.

Finally, combining Eqs.~(\ref{S28}) and (\ref{S59}), we obtain
\begin{align}
    \braket{\mS|\psi(t)}
   &=\me^{-\mi \omega_{\mS}t}\sum_{\ell=0}^{\infty} \qty(-g^{2})^\ell
   \qty[ \frac{S_D{\omega_0}^{-\nu} }{n}\me^{-\mi\nu\pi/2}\Gamma(\nu)\Gamma(1-\nu)]^\ell \frac{{}_1F_1(\ell-\ell\nu,2\ell-\ell\nu+1,\mi \omega_{\mS}t)}{\Gamma(2\ell-\ell\nu+1)} t^{2\ell-\ell\nu}\\
   &=\me^{-\mi \omega_{\mS}t}\sum_{\ell=0}^{\infty} \qty[-\frac{S_D\omega_0^{-\nu}\Gamma(\nu)\Gamma(1-\nu)}{n}\me^{-\mi\nu\pi/2}g^{2}t^{2-\nu}]^\ell \frac{1}{\Gamma(2\ell-\ell\nu+1)}
   \sum_{m=0}^{\infty}\frac{(\ell-\ell\nu)_m (\mi \omega_{\mS}t)^m}{(2\ell-\ell\nu+1)_m m!},
\end{align}
where we have used the expansion of the confluent hypergeometric function of the first kind given by
\begin{gather}
    {}_1F_1(a,b,z)=\sum_{m=0}^\infty \frac{(a)_m}{(b)_m}\frac{z^m}{m!},\\
    (a)_m:=\prod_{k=0}^{m-1}(a+k),~~~(a)_0=1.
\end{gather}
This completes the derivation of Eq.~(13) in the main text.
\section{B. Derivation of Eq.~(23)}
We here derive the survival amplitude with a finite cutoff in the form of Eq.~(23) in the main text.
\subsection{B.1 Second-order Dyson series expansion with a finite cutoff}
    We now consider the total Hamiltonian in the first quantization form with a cutoff $\Lambda$:
    \begin{align}
        H =& \omega_{\mathrm{S}} \ketbra{\mathrm{S}}{\mathrm{S}} +\omega_0 \int_{|\vec{k}|\leq \Lambda} |\vec{k}|^{n} \ketbra*{\vec{k}}\notag \\
        &+ g \int_{|\vec{k}|\leq \Lambda}\dd{\vec{k}} (\ketbra*{\mathrm{S}}{\vec{k}}  + \ketbra*{\vec{k}}{\mathrm{S}} ).\label{15}
      \end{align}
Note the change from Eq.~(8) in the main text, or Eq.~(\ref{S1}); the integration range is now finite. Thus, the second order of the Dyson series expansion (\ref{S2}) becomes
	\begin{align}
		 \braket{\mS|\psi(t)}&\simeq  \me^{-\mi \omega_{\mS}t}\qty[1- \int_{0}^{t} \dd{t_1} \int_{0}^{t_1} \dd{t_2} \braket{\mS|H_{\mathrm{int}}^{(\rm I)}(t_1)H_{\mathrm{int}}^{(\rm I)}(t_2)|\mS}]\\
         &= \me^{-\mi \omega_{\mS}t}\qty[1- g^2\int_{0}^{t} \dd{t_1} \int_{0}^{t_1} \dd{t_2} \me^{\mi \omega_{\mS}(t_1-t_2)}\int_{|\vec{k}|\leq \Lambda}\dd{\vec{k}} \me^{-\mi \omega_0|\vec{k}|^n  (t_1-t_2)}],
    \end{align}
    where
    $
    H_{\mathrm{int}}^{(\rm I)}(t) := \me^{\mi H_0 t} H_{\mathrm{int}} \me^{-\mi H_0 t}
    $
    is the coupling Hamiltonian in the interaction picture.
    
    We again change the variable to the polar coordinates and focus on the following double integral:
    \begin{align}
        \text{(The double integral)}&=\int_{0}^{t} \dd{t_1} \int_{0}^{t_1} \dd{t_2} \me^{\mi \omega_{\mS}(t_1-t_2)}\int_0^\Lambda \dd{k}  S_D  k^{D-1} \me^{-\mi \omega_0k^{n}  (t_1-t_2)}\label{S2nd}\\
        &=S_D \int_{0}^{t}\dd{t_1}\int_{0}^{t_1}\dd{\tau}\me^{\mi \omega_{\mS}\tau}\int_{0}^{\Lambda}\dd{k}k^{D-1}\me^{-\mi \omega_0 k^n  \tau},\label{S3}
	\end{align}
	where we set $t_1-t_2=\tau$. Let us further focus on the integral in Eq.~(\ref{S3}): 
    \begin{gather}
        f(t):=\int_{0}^{t}\dd{t_1}\int_{0}^{t_1}\dd{\tau}\me^{\mi \omega_{\mS}\tau}\int_{0}^{\Lambda}\dd{k}k^{D-1}\me^{-\mi \omega_0 k^n  \tau}.
    \end{gather} 
    Since the convergence radius is infinity from the d'Alembert's ratio test, we expand $\me^{-\mi \omega_0 k^n  \tau}$ and carry out the term-by-term integration as follows:
	\begin{align}
 f(t)&
		=\int_{0}^{t}\dd{t_1}\int_{0}^{t_1}\dd{\tau}\me^{\mi\omega_{\rm S}\tau}
		\sum_{\ell=0}^{\infty}
		\int_{0}^{\Lambda}\dd{k}
		\frac{(-\mi\omega_0 \tau)^\ell}{\ell !}
		k^{D+\ell n-1}\\
		&=\int_{0}^{t}\dd{t_1}\int_{0}^{t_1}\dd{\tau}\me^{\mi\omega_{\rm S}\tau}
		\Lambda^D\sum_{\ell=0}^{\infty}
		\frac{(-\mi\omega_0 \Lambda^{n} \tau)^\ell}{\ell!(D+\ell n)}.
    \end{align}
    Expanding the factor $\me^{\mi\omega_{\rm S}\tau}$ further, we have
    \begin{align}
		f(t)&=\Lambda^D\int_{0}^{t}\dd{t_1}
		\sum_{m=0}^{\infty}\sum_{\ell=0}^{\infty}
		\int_{0}^{t_1}\dd{\tau}
		\frac{(-\omega_0\Lambda^n)^\ell \mi^{\ell+m}\omega_{\rm S}^{m}}{\ell ! m! n(\nu+\ell)}\tau^{\ell+m}\\
		&=\Lambda^D\int_{0}^{t}\dd{t_1}
		\sum_{m=0}^{\infty}\sum_{\ell=0}^{\infty}
		\frac{(-\omega_0\Lambda^n)^\ell \mi^{\ell+m}\omega_{\rm S}^{m}}{\ell ! m! n(\nu+\ell)(\ell+m+1)}t_{1}^{\ell+m+1}\\
		&=\Lambda^D
		\sum_{m=0}^{\infty}\sum_{\ell=0}^{\infty}
		\frac{(-\omega_0\Lambda^n)^\ell \mi^{\ell+m}\omega_{\rm S}^{m}}{\ell ! m! n(\nu+\ell)(\ell+m+1)(\ell+m+2)}t^{\ell+m+2}.
	\end{align}
	Introducing a new variable $p:=\ell +m$, we can deform the summation as follows:
	\begin{align}
		f(t)&= \Lambda^D
		\sum_{p=0}^{\infty}\sum_{\ell=0}^{p}
		\frac{(-\omega_0\Lambda^n)^\ell \mi^{p}\omega_{\rm S}^{p-\ell}}{\ell ! (p-\ell)! n(\nu+\ell)(p+1)(p+2)}t^{p+2}.\label{7}
	\end{align}
	\subsection{B.2 The quadratic decay in the short-time limit}
	In the short-time regime $t\ll (\omega_0\Lambda^n)^{-1}$, the dominant term is given by $\ell=p=0$, which yields
	\begin{align}
        f(t)&\simeq
		 \frac{\Lambda^{D}}{2D}t^{2}.
	\end{align}
	Thus the second-order survival amplitude in the Dyson series expansion in this short-time limit is given as follows:
    \begin{align}
        \braket{\mS|\psi(t)}&\simeq  \me^{-\mi \omega_{\mS}t}\Bigg[1- g^2\int_{0}^{t} \dd{t_1} \int_{0}^{t_1} \dd{t_2} \me^{\mi \omega_{\mS}(t_1-t_2)}\int_{|\vec{k}|\leq \Lambda}\dd{\vec{k}} \me^{-\mi \omega_0|\vec{k}|^n (t_1-t_2)}\Bigg]\\
        &\simeq \me^{-\mi \omega_{\mS}t}\qty(1- g^2S_D \frac{\Lambda^{D}}{2D}t^{2}),
    \end{align}
which is the well-known quadratic decay \cite{facchi2008quantum}. We stress that this appears whenever there is an upper bound in the dispersion relation in addition to a lower bound.
\subsection{B.3 Emergence of the fractional power-law decay for large $\Lambda$}
	From now, we assume that $\omega_0\Lambda^n$ is much larger than $\omega_{\mS}$, which means that the time scale of the environment $1/(\omega_0\Lambda^n)$ is much shorter than that of the two-level system $1/\omega_{\mS}$. Thus in Eq.~(\ref{7}), the zeroth term of $\omega_{\mS}$ is dominant, which is specified by $\ell=p\geq 0$. Hence, Eq.~(\ref{7}) is reduced to
	\begin{align}
        f(t)&\simeq \Lambda^D
		 \sum_{p=0}^{\infty}
		 \dfrac{\mi^p (-\omega_0\Lambda^{n})^p}{p ! n(\nu+p)(p+1)(p+2)}t^{p+2}.\label{S54}
	\end{align}
	
	The goal is to prove the following equation:
\begin{align}
 (\text{r.h.s. of \ref{S54}})
         &=\omega_0^{-\nu}t^{2-\nu}\Biggl[-\dfrac{\mi^{-\nu}\Gamma(\nu-1)}{n(2-\nu)}-\dfrac{\mi (\omega_0\Lambda^{n}t)^{\nu-1}}{n(1-\nu)}+\mathcal{O}\qty((\omega_0\Lambda^n t)^{\nu-2})+\me^{-\mi\omega_0\Lambda^nt}\mathcal{O}\qty((\omega_0\Lambda^n t)^{\nu-3})\Biggr].\label{S12}
	\end{align}
	 The decomposition (\ref{S12}) implies the emergence of the fractional power-law decay in the limit of $\Lambda\to \infty$ as
	\begin{align}
		\lim\limits_{\Lambda\to\infty}(\text{r.h.s. of \ref{S54}})
		=-\dfrac{\mi^{-\nu}\omega_0^{-\nu}\Gamma(\nu-1)}{n(2-\nu)}t^{2-\nu}.
	\end{align} 
	
	Carrying out the partial-fraction decomposition in Eq.~(\ref{S54}), we obtain the following equation
	\begin{align}
        \Lambda^D
        \sum_{p=0}^{\infty}
        \dfrac{\mi^p (-\omega_0\Lambda^{n})^p}{p ! n(\nu+p)(p+1)(p+2)}t^{p+2}
		 &=
		  \Lambda^Dt^{2}
		  \sum_{p=0}^{\infty}
		  \dfrac{(-\mi\omega_0 \Lambda^{n} t)^{p}}{n(2-\nu)}\left[
		  \dfrac{1}{(p+1)!(p+\nu)}-\dfrac{1}{(p+2)!}
		  \right].\label{S14}
	\end{align}
	We can deform the first term on the right-hand side of Eq.~(\ref{S14}) as follows:
	\begin{align}
		\dfrac{\Lambda^Dt^{2}}{n(2-\nu)}
		 \sum_{p=0}^{\infty}
		 \dfrac{(-\mi\omega_0 \Lambda^{n} t)^{p}}{(p+1)!(p+\nu)}&=
		  \dfrac{\Lambda^Dt^{2}}{n(2-\nu)}
		  \sum_{q=1}^{\infty}\dfrac{(-\mi \omega_0\Lambda^{n} t)^{q-1}}{q!(q+\nu-1)}\label{Slhs}\\
		  &=
		  \dfrac{\Lambda^Dt^{2}}{n(2-\nu)}
		  (-\mi \omega_0\Lambda^{n} t)^{-1}
		  \sum_{q=1}^{\infty}\dfrac{(-\mi \omega_0\Lambda^{n} t)^{q}}{q!(q+\nu-1)}\\
		  &=
		  -\dfrac{\Lambda^Dt^{2}}{n(2-\nu)}
		  (\mi\omega_0\Lambda^{n}t)^{-\nu}
		  (\mi\omega_0\Lambda^{n}t)^{\nu-1}
		  \sum_{q=1}^{\infty}\dfrac{(-\mi \omega_0\Lambda^{n} t)^{q}}{q!(q+\nu-1)}\\
		  &=-\dfrac{\Lambda^Dt^{2}}{n(2-\nu)}
		  (\mi\omega_0\Lambda^{n}t)^{-\nu}
		  (\mi\omega_0\Lambda^{n}t)^{\nu-1}
		  \sum_{q=0}
		  ^{\infty}\dfrac{(-\mi \omega_0\Lambda^{n} t)^{q}}{q!(q+\nu-1)}\notag\\
		  &~~~~+\dfrac{\Lambda^Dt^{2}}{n(2-\nu)}
		  \dfrac{(\mi\omega_0\Lambda^{n}t)^{-1}}{\nu-1}.
    \end{align}
    Then we use the expansion of the upper incomplete gamma function \cite{gradshteyn2014table} given by
	\begin{align}
		\Gamma(a, x)&=\Gamma(a)-x^{a}\sum_{q=0}^{\infty}\dfrac{(-x)^{q}}{q!(q+a)}
	\end{align}
    for $a\neq 0,-1,-2,\ldots$,
	 where $\Gamma(a,x)$ is originally defined for $\mathrm{Re}~a>0$ and $|\arg x|<\pi$ as
	$
		 \Gamma(a,x)=\int_{x}^{\infty}\dd{\xi}\xi^{a-1}\me^{-\xi},
	$ and defined beyond the basic integral domain by analytic continuation. We note that we fix the principal branch of the complex logarithm $\log x$ with the branch cut on $\left(-\infty,0\right]$. Hence $x$ poses no branch issue.
    Thus Eq.~(\ref{Slhs}) is summarized as
    \begin{align}
        (\text{l.h.s. of \ref{Slhs}})&=-\dfrac{\Lambda^Dt^{2}}{n(2-\nu)}
		  (\mi\omega_0\Lambda^{n}t)^{-\nu}
		  \qty[\Gamma(\nu-1)-\Gamma(\nu-1,\mi\omega_0\Lambda^{n}t)]
		  +\dfrac{\Lambda^Dt^{2}}{n(2-\nu)(\nu-1)}{(\mi\omega_0\Lambda^{n}t)^{-1}}\\
		  &=
		  -\dfrac{\mi^{-\nu}\omega_0^{-\nu}t^{2-\nu}}{n(2-\nu)}\qty[\Gamma(\nu-1)-\Gamma(\nu-1,\mi\omega_0\Lambda^{n}t)]
		  +\dfrac{\mi\Lambda^{D-n}t}{n(2-\nu)(1-\nu)\omega_0},
	\end{align}
    where we have used the fact $D-n\nu=0$.
     From the asymptotic series of the upper incomplete gamma function
     \begin{gather}
        \Gamma(a,x)\sim x^{a-1}\me^{-x}\sum_{k=0}\frac{\Gamma(a)}{\Gamma(a-k)}x^{-k}~~~\text{in the limit}~ |x|\to \infty ~~\text{and}~~ |\arg x|<\frac{3\pi}{2},
     \end{gather}
     we have
    \begin{align}
        \Gamma(\nu -1,\mi\omega_0\Lambda^n t)= \qty(\mi \omega_0\Lambda^nt)^{\nu-2}\me^{-\mi\omega_0\Lambda^nt}\qty[1+\mathcal{O}\qty((\omega_0\Lambda^n t)^{-1})].
    \end{align}
    Therefore, Eq.~(\ref{Slhs}) is reduced to
    \begin{align}
        (\text{l.h.s. of \ref{Slhs}})&
         =-\dfrac{\mi^{-\nu}\omega_0^{-\nu}t^{2-\nu}}{n(2-\nu)}\qty[\Gamma(\nu-1)-\Gamma(\nu-1,\mi\omega_0\Lambda^{n}t)]
         +\dfrac{\mi\Lambda^{D-n}t}{n(2-\nu)(1-\nu)\omega_0}
         \\
         &=-\dfrac{\mi^{-\nu}\omega_0^{-\nu}t^{2-\nu}}{n(2-\nu)}\qty{\Gamma(\nu-1)-\qty(\mi \omega_0\Lambda^nt)^{\nu-2}\me^{-\mi\omega_0\Lambda^nt}\qty[1+\mathcal{O}\qty((\omega_0\Lambda^n t)^{-1})]}\notag\\
		  &~~~+\dfrac{\mi\Lambda^{D-n}t}{n(2-\nu)(1-\nu)\omega_0}.\label{S96}
    \end{align}

	For the second term on the right-hand side of Eq.~(\ref{S14}), we can deform it as follows:
	\begin{align}
		 \dfrac{\Lambda^Dt^{2}}{n(2-\nu)}
		 \sum_{p=0}^{\infty}\dfrac{(-\mi\omega_0\Lambda^{n}t)^{p}}{(p+2)!}&=\dfrac{\Lambda^Dt^{2}}{n(2-\nu)}
		 \sum_{q=2}^{\infty}\dfrac{(-\mi\omega_0\Lambda^{n}t)^{q-2}}{q!}\\
		 &=
		 \dfrac{\Lambda^Dt^{2}}{n(2-\nu)}
		 (-\mi\omega_0\Lambda^{n}t)^{-2}
		 \sum_{q=2}^{\infty}\dfrac{(-\mi\omega_0\Lambda^{n}t)^{q}}{q!}\\
		 &=-
		 \dfrac{\Lambda^{D-2n}}{n(2-\nu)\omega_0^2}
		 \qty[
		 \me^{-\mi\omega_0\Lambda^{n}t}-1-(-\mi\omega_0\Lambda^{n}t)
		 ]\\
		 &=-
		 \dfrac{\Lambda^{D-2n}}{n(2-\nu)\omega_0^2}
		 \qty(
		 \me^{-\mi\omega_0\Lambda^{n}t}-1
		 )
		 -\dfrac{\mi \Lambda^{D-n}t}{n(2-\nu)\omega_0}.\label{S100}
	\end{align}
	Combining Eqs.~(\ref{S96}) and (\ref{S100}) expressing the first and second terms of Eq.~(\ref{S14}), we obtain the following decomposition:
		\begin{align}
            f(t)
		&\simeq\Lambda^D
        \sum_{p=0}^{\infty}
        \dfrac{\mi^p (-\omega_0\Lambda^{n})^p}{p ! n(\nu+p)(p+1)(p+2)}t^{p+2}\\
        &=-\dfrac{\mi^{-\nu}\omega_0^{-\nu}t^{2-\nu}}{n(2-\nu)}\qty{\Gamma(\nu-1)-\qty(\mi \omega_0\Lambda^nt)^{\nu-2}\me^{-\mi\omega_0\Lambda^nt}\qty[1+\mathcal{O}\qty((\omega_0\Lambda^n t)^{-1})]}\notag\\
        &~~~~
        +\dfrac{\mi\Lambda^{D-n}t}{n(2-\nu)(1-\nu)\omega_0}+
		\dfrac{\Lambda^{D-2n}}{n(2-\nu)\omega_0^2}
		\qty(
		\me^{-\mi\omega_0\Lambda^{n}t}-1
		)
		+\dfrac{\mi \Lambda^{D-n}t}{n(2-\nu)\omega_0}.\label{S101}
        \end{align}
        Combining the second and fourth terms on the right-hand side of the above expression yields the following equation:
        \begin{align}
            \text{(r.h.s. of (\ref{S101}))}&=-\dfrac{\mi^{-\nu}\omega_0^{-\nu}t^{2-\nu}}{n(2-\nu)}\qty{\Gamma(\nu-1)-\qty(\mi \omega_0\Lambda^nt)^{\nu-2}\me^{-\mi\omega_0\Lambda^nt}\qty[1+\mathcal{O}\qty((\omega_0\Lambda^n t)^{-1})]}\notag\\
        &~~~~
		+\dfrac{\mi\Lambda^{D-n}t}{n(1-\nu)\omega_0}+
		\dfrac{\Lambda^{D-2n}}{n(2-\nu)\omega_0^2}
		\qty(
		\me^{-\mi\omega_0\Lambda^{n}t}-1
        ).\label{S102}
        \end{align}
By factoring out $\omega_0^{-\nu}t^{2-\nu}$, we recast the above expression into the sum of dimensionless quantities as follows:
        \begin{align}
            \text{(r.h.s. of (\ref{S101}))}&=-\dfrac{\mi^{-\nu}\omega_0^{-\nu}t^{2-\nu}}{n(2-\nu)}\qty{\Gamma(\nu-1)-\qty(\mi \omega_0\Lambda^nt)^{\nu-2}\me^{-\mi\omega_0\Lambda^nt}\qty[1+\mathcal{O}\qty((\omega_0\Lambda^n t)^{-1})]}\notag\\
        &~~~~
		+\omega_0^{-\nu}t^{2-\nu}\dfrac{\mi (\omega_0\Lambda^{n}t)^{\nu-1}}{n(1-\nu)}+
		\omega_0^{-\nu}t^{2-\nu}\dfrac{(\omega_0\Lambda^n t)^
        {\nu-2}}{n(2-\nu)}
		\qty(
		\me^{-\mi\omega_0\Lambda^{n}t}-1
        )\\
        &=\omega_0^{-\nu}t^{2-\nu}\Biggl[-\dfrac{\mi^{-\nu}\Gamma(\nu-1)}{n(2-\nu)}-\frac{\qty(\omega_0\Lambda^nt)^{\nu-2}}{n(2-\nu)}\me^{-\mi\omega_0\Lambda^nt}\qty[1+\mathcal{O}\qty((\omega_0\Lambda^n t)^{-1})]\notag\\
        &~~~~+\dfrac{\mi (\omega_0\Lambda^{n}t)^{\nu-1}}{n(1-\nu)}+\dfrac{(\omega_0\Lambda^n t)^
        {\nu-2}}{n(2-\nu)}
		\qty(
		\me^{-\mi\omega_0\Lambda^{n}t}-1
        )\Biggr].\label{S105}
    \end{align}
    Finally, combining the second and fourth terms in the square brackets of the above expression, we have
    \begin{align}
        \text{(r.h.s. of (\ref{S101}))}&=\omega_0^{-\nu}t^{2-\nu}\Biggl[-\dfrac{\mi^{-\nu}\Gamma(\nu-1)}{n(2-\nu)}+\dfrac{\mi (\omega_0\Lambda^{n}t)^{\nu-1}}{n(1-\nu)}+\mathcal{O}\qty((\omega_0\Lambda^n t)^{\nu-2})+\me^{-\mi\omega_0\Lambda^nt}\mathcal{O}\qty((\omega_0\Lambda^n t)^{\nu-3})\Biggr].
	\end{align}
    We thus finally arrive at the following expression:
    \begin{gather}
        f(t)\simeq \omega_0^{-\nu}t^{2-\nu}\Biggl[-\dfrac{\mi^{-\nu}\Gamma(\nu-1)}{n(2-\nu)}+\dfrac{\mi (\omega_0\Lambda^{n}t)^{\nu-1}}{n(1-\nu)}+\mathcal{O}\qty((\omega_0\Lambda^n t)^{\nu-2})+\me^{-\mi\omega_0\Lambda^nt}\mathcal{O}\qty((\omega_0\Lambda^n t)^{\nu-3})\Biggr].
    \end{gather}

    Then the survival amplitude is given by:
    \begin{align}
            \braket{\mS|\psi(t)}\simeq  \me^{-\mi \omega_{\mS}t}&\Bigg\{1- g^2S_D \omega_0^{-\nu}t^{2-\nu}\Biggl[-\dfrac{\mi^{-\nu}\Gamma(\nu-1)}{n(2-\nu)}+\dfrac{\mi (\omega_0\Lambda^{n}t)^{\nu-1}}{n(1-\nu)}\notag\\
            &~~+\mathcal{O}\qty((\omega_0\Lambda^n t)^{\nu-2})+\me^{-\mi\omega_0\Lambda^nt}\mathcal{O}\qty((\omega_0\Lambda^n t)^{\nu-3})\Biggr]\Bigg\}\\
            =\me^{-\mi \omega_{\mS}t}&\Bigg\{1- g^2S_D \omega_0^{-\nu}t^{2-\nu}\Biggl[-\dfrac{\mi^{-\nu}\Gamma(\nu-1)}{n(2-\nu)}+\mathcal{O}\qty((\omega_0\Lambda^n t)^{\nu-1})\Biggr]\Bigg\}.
    \end{align}
    This completes the derivation of Eq.~(23).
	Since the power with respect to $\omega_0\Lambda^n t$ is negative and only the normal gamma function remains for $\omega_0\Lambda^n\to\infty$, we can finally see the fractional power in the second-order contribution in the limit of $\Lambda\to\infty$ as follows:
    \begin{align}
        \lim_{\Lambda\to\infty}\braket{\mS|\psi(t)}&\simeq  \me^{-\mi \omega_{\mS}t}\Bigg[1+ \dfrac{\mi^{-\nu}g^2S_D \omega_0^{-\nu}\Gamma(\nu-1)}{n(2-\nu)}t^{2-\nu}\Bigg]\\
        &=\me^{-\mi \omega_{\mS}t}\qty[1-\frac{\nu \Gamma(\nu)\pi^{D/2}\me^{-\mi \nu \pi/2}}{\Gamma(\frac{D}{2}+1)(1-\nu)(2-\nu)}
        g^2{\omega_0}^{-\nu}t^{2-\nu}],
\end{align}
which is the same as Eq.~(14) in the main text. Here we have used $S_D={2\pi^{D/2}/\Gamma(D/2)}$, $\nu=D/n$ and the shifted factorial property of the gamma function $\Gamma(x+1)=x\Gamma(x)$ for all complex $x$ except for integers less than or equal to zero. 
\section{C. Detailed calculation of dynamics in the long-time limit}
We here show the detailed calculation written in the section of \textit{Dynamics in the long-time limit} in the main text.
\subsection{C.1 Derivation of Eq.~(24)}
First, we derive the expression in the Feshbach formalism \cite{feshbach1958unified,hatano2014time}, Eq.~(24) in the main text. Let us first express the survival amplitude as follows:
\begin{align}
    \braket{\mathrm{S}|\psi(t)}&:= \bra{\mS}\me^{-\mi H t}\ket{\mS}\\
    &=\int \dd{E} \bra{\mS}\me^{-\mi H t}\ket{E}\braket{E|\mS}\\
    &=\int \dd{E} \me^{-\mi E t}\braket{\mS|E}\braket{E|\mS}.\label{S103}
\end{align}
Here, $\ket{E}$ is an energy eigenstate with an eigenvalue $E$. 
We now compute the Fourier transform of Eq.~(\ref{S103}), but the integral may fail to converge.
Since we are interested in the initial-value problem ($t>0$), in order to avoid the convergence problem at large time, we introduce the function
\begin{gather}
    G(t):=\theta(t)\braket{\mathrm{S}|\psi(t)}\me^{-\eta t},
\end{gather}
where $\theta(t)$ is the Heaviside step function and $\eta$ is a positive infinitesimal constant.

Next, we perform the Fourier transform of $G(t)$:
\begin{align}
    \Tilde{G}(x)&:=\int_{-\infty}^{\infty}\dd{t} G(t)\me^{\mi xt}\\
    &=\int_{0}^{\infty}\dd{t} \braket{\mathrm{S}|\psi(t)}\me^{\mi xt}\me^{-\eta t}\\
    &=\int_{0}^{\infty}\dd{t} \int \dd{E}\abs{\braket{\mS|E}}^2\me^{\mi(x+\mi \eta-E)t }\\
    &=\int \dd{E} \frac{\mi}{x+\mi \eta -E}\abs{\braket{\mS|E}}^2.
\end{align} 
We then perform the inverse Fourier transform:
\begin{align}
    G(t)&=\frac{1}{2\pi}\int_{-\infty}^{\infty}\dd{x} \Tilde{G}(x)\me^{-\mi x t}\\
    &=\frac{1}{2\pi}\bra{\mS} \qty(\int_{-\infty}^{\infty}\dd{x} \int \dd{E} \frac{\mi \me^{-\mi x t}}{x+\mi \eta -E})\ket{E}\braket{E|\mS}\\
    &=\frac{1}{2\pi}\bra{\mS}\qty( \int_{-\infty+\mi \eta}^{\infty+\mi \eta}\dd{z} \int \dd{E} \frac{\mi \me^{-\mi z t}}{z -E})\ket{E}\braket{E|\mS}\me^{-\eta t}\label{S84}\\
    &=\frac{1}{2\pi \mi}\bra{\mS} \qty( \int_{\infty+\mi \eta}^{-\infty+\mi \eta}\dd{z} \frac{\me^{-\mi z t}}{z -H})\ket{\mS}\me^{-\eta t}\label{S85}.
\end{align}
From Eq.~(\ref{S84}) to Eq.~(\ref{S85}), we have used the following relation:
\begin{gather}
    \int \dd{E} \frac{1}{z-E}\ketbra{E}=\frac{1}{z-H}.
\end{gather}
Thus, the survival amplitude for $t>0$ is expressed as
\begin{gather}
    \braket{\mathrm{S}|\psi(t)}=\frac{1}{2\pi \mi}\bra{\mS} \qty(\int_{\infty+\mi \eta}^{-\infty+\mi \eta}\dd{z} \frac{\me^{-\mi z t}}{z -H})\ket{\mS}.
\end{gather}
This is a retarded Green's function of the survival amplitude. Hereafter we use the projection operators $\mP$ and $\mQ$ defined as follows:
\begin{gather}
    \mP:=\ketbra{\mS},~\mQ:= I-\mP=\int_{\mathbb{R}^D}\dd \vec{k} \ketbra*{\vec{k}},\\
    \mP H\mP=H_{\mathrm{S}},~\mQ H\mQ=H_{\mathrm{E}}= \int_{\mathbb{R}^D}\dd \vec{k} \omega_0|\vec{k}|^n \ketbra*{\vec{k}}{\vec{k}},\\
    \mP H\mQ=g\int_{\mathbb{R}^D}\dd \vec{k}\ketbra*{\mS}{\vec{k}},~ \mQ H\mP=g\int_{\mathbb{R}^D}\dd \vec{k}\ketbra*{\vec{k}}{\mS}.
\end{gather}
Since the projection operator $\mP$ projects onto the state $\ketbra{\mS}$, we can insert $\mP$ as follows:
\begin{align}
    \braket{\mathrm{S}|\psi(t)}&=\frac{1}{2\pi \mi}\bra{\mS} \mP\qty( \int_{\infty+\mi \eta}^{-\infty+\mi \eta}\dd{z} \frac{\me^{-\mi z t}}{z -H})\mP\ket{\mS}.
\end{align}
Using the resolvent expansion \cite{hatano2014time}, we have
\begin{gather}
    \mP \frac{1}{z-H}\mP=\mP\frac{1}{z-H_{\mathrm{eff}}(z)}\mP,
\end{gather}
where the effective Hamiltonian is given by
\begin{align}
    H_{\mathrm{eff}}(z):= \mP H\mP +\mP H\mQ\frac{1}{z-\mQ H\mQ}\mQ H\mP.\label{effec}
  \end{align}
Therefore, the expression for the survival amplitude becomes
\begin{align}
    \braket{\mathrm{S}|\psi(t)}&=\frac{1}{2\pi \mi}\bra{\mS} \mP\qty( \int_{\infty+\mi \eta}^{-\infty+\mi \eta}\dd{z} \frac{\me^{-\mi z t}}{z -H_{\mathrm{eff}}(z)})\mP\ket{\mS}\\
    &=\frac{1}{2\pi \mi} \int_{\infty+\mi \eta}^{-\infty+\mi \eta}\dd{z} \frac{\me^{-\mi z t}}{z -\bra{\mS}H_{\mathrm{eff}}(z)\ket{\mS}}.\label{199}
\end{align}
This completes the derivation of Eq.~(24) in the main text.
\subsection{C.2 Detailed calculation of the effective Hamiltonian (25)}
Next, we show that the effective Hamiltonian (25) has the form (30) for $0<\nu<1$ in the main text. We transform Eq.~(\ref{effec}) as follows:
\begin{align}
    H_{\mathrm{eff}}(z)
    &=\omega_{\mS}\ketbra{\mS}+g^2\int_{\mathbb{R}^D} \dd{\vec{k}} \ketbra*{\mS}{\vec{k}} \frac{1}{z-H_{\mathrm{E}}} \int_{\mathbb{R}^D} \dd{\vec{k}'} \ketbra*{\vec{k}'}{\mS}\\
    &=\omega_{\mS}\ketbra{\mS}+g^2 \int_{\mathbb{R}^D} \dd{\vec{k}}\frac{1}{z-\omega_0|\vec{k}|^n}\ketbra{\mS}.\label{S60}
\end{align}
Let us focus on evaluating the integral in the second term on the right-hand side of Eq.~(\ref{S60}):
\begin{align}
    \int_{\mathbb{R}^D} \dd{\vec{k}}\frac{1}{z-\omega_0|\vec{k}|^{n}}
    &=-\int_0^\infty \dd{k} S_D \frac{k^{D-1}}{\omega_0k^n -z}\\
    &=-\frac{S_D \omega_0^{-\nu}}{n}\int_0^\infty \dd{\xi}\frac{\xi^{\nu-1}}{\xi-z},\label{S63}
\end{align}
where we have changed the variable to the polar coordinate and set $\omega_0k^n =\xi$. 
For evaluation of the integral in Eq.~(\ref{S63}), we use the Cauchy residue theorem, taking the contour as depicted in Fig.~S1.
    \begin{figure}[h]
            \centering
            \includegraphics[width=0.5\columnwidth]{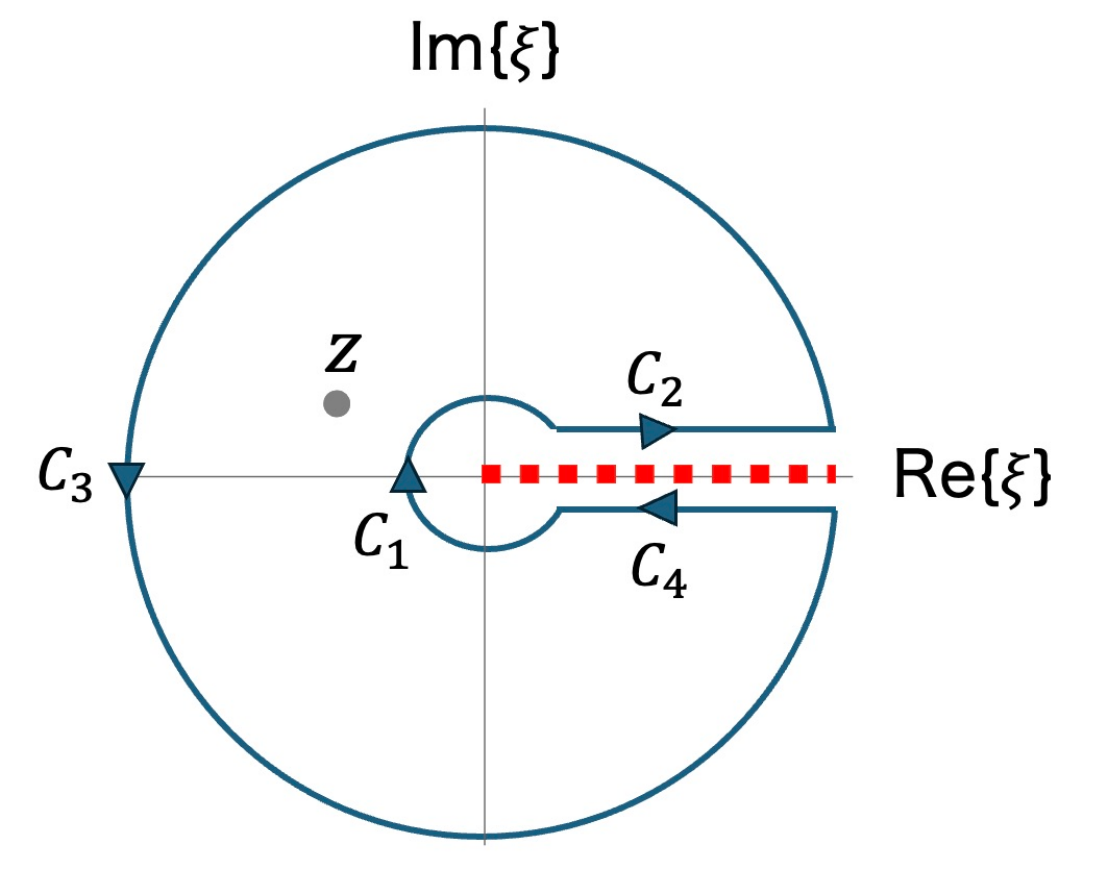}
            \caption{Contour used for the calculation of the integral of Eq.~(\ref{S63}). The path $C_1$ represents a small circular arc of radius $\epsilon$ (with $\epsilon \to 0$). The path $C_2$ extends from $\epsilon$ to $\infty$ and the path $C_4$ extends from $\infty \me^{\mi 2\pi} $ to $\epsilon\me^{\mi 2\pi}$. The path $C_3$ represents a large circular arc of radius $R$ (with $R \to \infty$). The point $z$ denotes a singular point of the integrand.}
        \end{figure}

        Let us first evaluate the integral of each path.
             The integral along $C_1$ is given as follows:
            \begin{align}
                \abs{\int_{C_1}\dd{\xi}\frac{\xi^{\nu-1}}{\xi-z}}&=\abs{\mi \epsilon \int_{2\pi}^0\dd{\theta} \me^{\mi \theta} \frac{(\epsilon \me^{\mi \theta})^{\nu-1}}{\epsilon \me^{\mi \theta}-z}}\\
                &= \abs{\int_{2\pi}^0\dd{\theta}  \frac{\epsilon^\nu (\me^{\mi \theta})^{\nu}}{\epsilon \me^{\mi \theta}-z}}\to 0 \quad\text{as}\quad \epsilon \to 0.
            \end{align}
             The integrals along $C_2$ and $C_4$ are summarized as follows:
        \begin{align}
            \int_{C_2+C_4}\dd{\xi}\frac{\xi^{\nu-1}}{\xi-z}&=\int_0^\infty \dd{\xi}\frac{\xi^{\nu-1}}{\xi-z}+\int_{\infty}^0 \dd{\xi}\frac{\xi^{\nu-1}\me^{\mi2(\nu-1)\pi}}{\xi \me^{\mi2\pi}-z}\\
            &=\int_0^\infty \dd{\xi}\frac{\xi^{\nu-1}}{\xi-z}-\me^{\mi2(\nu-1)\pi}\int_0^{\infty} \dd{\xi}\frac{\xi^{\nu-1}}{\xi -z}\\
            &=(1-\me^{\mi2(\nu-1)\pi})\int_0^\infty \dd{\xi}\frac{\xi^{\nu-1}}{\xi-z}.\label{S141}
        \end{align}
         The integral along $C_3$ is given as follows:
        \begin{align}
            \abs{\int_{C_3}\dd{\xi}\frac{\xi^{\nu-1}}{\xi-z}}&=\abs{\mi R \int_0^{2\pi}\dd{\theta} \me^{\mi \theta} \frac{(R \me^{\mi \theta})^{\nu-1}}{R \me^{\mi \theta}-z}}\\
            &= \abs{\int_0^{2\pi}\dd{\theta}  \frac{R^\nu (\me^{\mi \theta})^{\nu}}{R \me^{\mi \theta}-z}}\to 0 ~~\text{as}~~R\to\infty
        \end{align}
        according to Jordan's lemma. Therefore, only the remaining integral is Eq.~(\ref{S141}).
        On the other hand, the residue theorem gives the contribution of the residue at $\xi=z$:
        \begin{gather}
            \int_{C_1+C_2+C_3+C_4}\dd{\xi}\frac{\xi^{\nu-1}}{\xi-z}
            =2\pi\mi z^{\nu-1}.\label{S144}
        \end{gather}
        
        Equating Eqs.~(\ref{S141}) and (\ref{S144}), we obtain
        \begin{align}
            \int_0^\infty \dd{\xi}\frac{\xi^{\nu-1}}{\xi-z}&=\frac{2\pi\mi}{1-\me^{\mi2(\nu-1)\pi}} z^{\nu-1}\\
            &=\frac{\pi}{\sin[(1-\nu)\pi]} \me^{\mi(1-\nu)\pi} z^{\nu-1}.\label{S75}
        \end{align}
        Inserting this result into Eq.~(\ref{S63}) and Eq.~(\ref{S60}), we find the effective Hamiltonian in the form
        \begin{gather}
            H_{\mathrm{eff}}(z)=\qty[\omega_{\mS}-g^2\frac{S_D\omega_0^{-\nu} }{n} \frac{\pi\me^{\mi(1-\nu)\pi}}{\sin[(1-\nu)\pi]} z^{\nu-1}]\ketbra{\mS}.
        \end{gather}

        Therefore, the survival amplitude (\ref{199}) is given as follows:
        \begin{gather}
            \braket{\mS|\psi(t)}=\frac{1}{2\pi \mi} \int_{\infty+\mi \eta}^{-\infty+\mi \eta}\dd{z} F(z),\label{S151}\\
             F(z):=\frac{\me^{-\mi z t}}
             {z -\omega_{\mS}+B \me^{\mi (1-\nu )\pi } z^{\nu-1}},\\
             B:=g^2 \frac{S_D\omega_0^{-\nu}}{n} \frac{\pi}{\sin[(1-\nu )\pi]},
           \end{gather}  
           which is the same as Eqs.~(31)--(33) in the main text.

In the Supplemental material C.3 and C.4 below, we will examine the poles of $F(z)$ and in the Supplemental material C.5 we will evaluate the branch-cut contribution to Eq.~(\ref{S151}).
\subsection{C.3 Finding the singular points of Eq.~(32)}
To evaluate the complex integral in Eq.~(\ref{S151}) based on the residue theorem, we analyze the singularities of $F(z)$. The singularities are solutions of the following equation from the denominator of $F(z)$:
\begin{gather}
    z=\omega_{\mS}-B \me^{\mi(1-\nu)\pi} z^{\nu-1}.\label{225}
\end{gather}
We parametrize the complex variable as $z = r \me^{\mi\theta}~~(r\ge0,\;\theta\in\mathbb{R}),$ and define the first Riemann sheet by
$
-\pi/2 \le \theta < 3\pi/2
~~\text{for}~~ 0<\nu<1
$
as is specified in Fig.~2 in the main text.
Equation~(\ref{225}) reduces to the imaginary part
   \begin{gather}
    r\sin\theta=-B r^{\nu-1}\sin \phi,\label{sin}
   \end{gather}
   and the real part
\begin{gather}
    r\cos\theta=\omega_{\mS}-B r^{\nu-1}\cos \phi\label{cos},
\end{gather}
where
$
    \phi:=(1-\nu)(\pi-\theta).
$
   Let us examine where the solutions exist.
   A necessary condition for the existence of the solutions is that the signs of $\sin\theta$ and $\sin\phi$ are opposite by Eq.~(\ref{sin}). Focusing on the sign of $\sin\theta$, we hereafter divide the discussion into the following regions.
\begin{itemize}
    \item Region $-\pi/2\le\theta<0$ (i.e. $\sin\theta<0$):\\
   The function $F(z)$ has at most one singularity in this region.
The conditions of $\mathrm{sign}(\sin\theta)=-\mathrm{sign}(\sin\phi)$ are
\begin{gather}
    -\frac{\pi}{2}<\theta<0~~\text{for }\nu>\frac{1}{3},\label{S145}\\
    -\frac{\nu}{1-\nu}\pi<\theta<0 ~~\text{for }\nu\leq \frac{1}{3}.\label{S146}
\end{gather}
From now, we restrict our discussion to the region above. Note that in this region, $\sin\theta<0$ and $\sin\phi>0$.
   Inserting Eq.~(\ref{sin}) into Eq.~(\ref{cos}), we obtain
   \begin{align}
    \omega_{\mS}&=B \qty(-\frac{B\sin\phi}{\sin\theta})^{\frac{\nu-1}{2-\nu}}\cos\phi +\qty(-\frac{B\sin\phi}{\sin\theta})^{\frac{1}{2-\nu}}\cos\theta\\
    &=\qty(-\frac{B\sin\phi}{\sin\theta})^{\frac{1}{2-\nu}}\qty[\cos\theta +\cos\phi \qty(-\frac{\sin\phi}{\sin\theta})^{-1}]\\
    &=\qty(-\frac{B\sin\phi}{\sin\theta})^{\frac{1}{2-\nu}}\frac{\sin(\phi-\theta)}{\sin\phi}=:R(\theta).\label{SR}
   \end{align}
Next, we prove that there exists at most one solution by the monotonicity of $R(\theta)$ in this region.
   Differentiating $ \log |R(\theta)|$ gives
   \begin{align}
    \frac{\dd \log \abs{R(\theta)}}{\dd \theta}&=\dv{R(\theta)}{\theta}\frac{1}{R(\theta)}\\
    &=\frac{1}{2-\nu}\qty[(\nu-1)\cot \phi-\cot \theta]+(\nu-2)\cot (\phi-\theta)-(\nu-1)\cot \phi\\
    &=\frac{(1-\nu)^2}{2-\nu}\cot \phi -\frac{1}{2-\nu}\cot \theta+(\nu-2)\frac{\cot \phi\cot \theta +1}{\cot \theta -\cot \phi}\\
    &=\frac{1}{(2-\nu)(\cot \theta -\cot \phi)}\Big[(1-\nu)^2(\cot \theta \cot \phi -\cot^2 \phi)-\cot^2 \theta +\cot \theta\cot \phi\notag\\
    &~~~~-(2-\nu)^2\cot \theta \cot \phi - (2-\nu)^2\Big]\\
    &=\frac{-1}{(2-\nu)(\cot \theta -\cot \phi)}\qty{\qty[(1-\nu)\cot \phi +\cot \theta]^2 + (2-\nu)^2}\\
    &=\frac{-\sin \theta \sin \phi}{(2-\nu)\sin(\phi-\theta)}\qty{\qty[(1-\nu)\cot \phi +\cot \theta]^2 + (2-\nu)^2}.
   \end{align}
   Thus, we have
   \begin{align}
    \dv{R(\theta)}{\theta}&=R(\theta)\frac{\dd \log \abs{R(\theta)}}{\dd \theta}\\
    &=-\frac{\sin\theta}{2-\nu}\qty(-\frac{B\sin\phi}{\sin\theta})^{\frac{1}{2-\nu}}\qty{\qty[(1-\nu)\cot \phi +\cot \theta]^2 + (2-\nu)^2}>0
   \end{align}
 
   Since the derivative is positive, $R(\theta)$ is monotonically increasing in both cases (\ref{S145}) and (\ref{S146}). Note that $R(\theta)$ is obviously a continuous function.
   Therefore, there is at most one solution to $\omega_{\mS}=R(\theta)$ with $-\pi/2\le\theta<0$. There exists only one solution if
   \begin{gather}
    R(-{\pi/2})<\omega_{\mS}<R(0)~~\text{for }\nu>\frac{1}{3},\\
    R({-\nu\pi/(1-\nu)})<\omega_{\mS}<R(0) ~~\text{for }\nu\leq \frac{1}{3}.
\end{gather}
Otherwise, Eq.~(\ref{225}) has no solutions.
   If such a solution exists, it corresponds to a pole of $F(z)$. Its residue only produces the exponential decay in the time evolution, and hence does not affect the long-time dynamics. We therefore neglect it in the present asymptotic analysis.
    \item Region $0\leq \theta <\pi$ (i.e. $\sin\theta\geq 0$):\\
    No singularities of $F(z)$ exist in this region. Any singularity has to satisfy Eq.~(\ref{225}) with the inequalities
      \begin{gather}
          0<\phi \leq (1-\nu)\pi,
      \end{gather}
      for which the left- and right-hand sides of Eq.~(\ref{sin}) have opposite signs. Hence no solutions to Eq.~(\ref{225}) exist.
    \item Line $\theta =\pi$ (i.e. $\sin\theta =0$):\\
    Exactly one singularity of $F(z)$ is present on the negative real axis, where Eq.~(\ref{225}) reduces to
      \begin{gather}
      -r=\omega_{\mS}-B r^{\nu-1}.\label{229}
      \end{gather}
      Geometrical consideration reveals that there is a unique intersection of the lines
$r+\omega_{\mS}$ and $B\,r^{\nu-1}$. We therefore have one negative solution to Eq.~(\ref{225}), which is denoted by $z_0$ in the main text.   
      
      \item Region $\pi< \theta <3\pi/2$ (i.e. $\sin\theta<0$):\\
     No singularities of $F(z)$ exist in this region. Equation (\ref{sin}) with the inequalities
      \begin{gather}
          -(1-\nu){\pi}/2<\phi<0
      \end{gather}
      implies that the left-hand side of Eq.~(\ref{sin}) is negative, whereas that of the right-hand side is positive.
Thus, no solutions exist in this region.
    \end{itemize}

To summarize, we need only one pole $z_0$ on the negative part of the real axis for our asymptotic analysis here.    
\subsection{C4. Proof that $F(z)$ has only simple poles}
In this section we prove that every pole of $F(z)$ is simple.  
Let us define the denominator of $F(z)$ as
\begin{gather}
    D(z)=z-\omega_{\mS}+B\me^{\mi(1-\nu)\pi}z^{\nu-1}.
\end{gather}
A pole of $F(z)$ is multiple if and only if there exists a point $\zeta\in \mathbb{C}$ such that $D(\zeta)=0$ and $D'(\zeta)=0$. Let us first assume that such a pole $\zeta$ exists and show that it leads to a contradiction. The two conditions read
\begin{gather}
    D(\zeta)=\zeta-\omega_\mS +B\me^{\mi(1-\nu)\pi}\zeta^{\nu-1}=0,\label{S125}\\
    D'(\zeta)=1+(\nu-1)B\me^{\mi(1-\nu)\pi}\zeta^{\nu-2}=0,\label{S126}
\end{gather}
which yield
\begin{gather}
    \zeta=\frac{1-\nu}{2-\nu}\omega_\mS.
\end{gather}
Substituting this result back into Eq.~(\ref{S126}), we find that the left-hand side acquires a non-vanishing imaginary part for real $\omega_\mS$ and $B$. This result is a contradiction. Therefore, no simultaneous solution of Eqs.~(\ref{S125}) and (\ref{S126}) exists, and we thus conclude that every pole of $F(z)$ is simple.
\subsection{C.5 Calculation of the asymptotic expansion (37)}
In order to calculate the complex integral (24) in the main text (or Eq.~(\ref{S151})), we have to evaluate not only the pole contribution but also the branch-cut one. In this section, we consider the latter. It comes from the integral $F(z)$ along the contours $C_a$ and $C_b$ of Fig.~3 in the main text. It follows that
\begin{align}
    \int _{C_\mathrm{a}+C_\mathrm{b}}\dd{z} F(z)
    &=\int_{\infty}^{0}\dd{(z\me^{\mi 3\pi/2})} \frac{\me^{-\mi z\me^{\mi 3\pi/2} t}}{z\me^{\mi 3\pi/2} -\qty(\omega_{\mS}-B\me^{\mi(1-\nu)\pi} (z\me^{\mi 3\pi/2})^{\nu-1})}\notag\\
    &~~~+ \int_{0}^{\infty}\dd{(z\me^{-\mi \pi/2})} \frac{\me^{-\mi z\me^{-\mi \pi/2} t}}{z\me^{-\mi \pi/2} -\qty(\omega_{\mS}-B\me^{\mi(1-\nu)\pi} (z\me^{-\mi \pi/2})^{\nu-1})}\\
      &=\mi \int_{0}^{\infty}\dd{z} \Bigg(\frac{\me^{- zt}}{-\mi z -\omega_{\mS}+B
   \me^{\mi(1-\nu)(-\pi/2)} z^{\nu-1}}-\frac{\me^{- zt}}{-\mi z -\omega_{\mS}+B
   \me^{\mi(1-\nu)(3\pi/2)} z^{\nu-1}}\Bigg).\label{53}
\end{align}
By setting $u=zt$, we obtain
\begin{align}
    \int _{C_\mathrm{a}+C_\mathrm{b}}\dd{z} F(z)
    &=\frac{\mi}{t } \int_{0}^{\infty}\dd{u} \Bigg(\frac{\me^{- u}}{-\mi u/t -\omega_{\mS}+B
   \me^{-\mi(1-\nu)\pi/2} (u/t)^{\nu-1}}\notag\\
   &~~~~~~~~~~~~~~~~~~-\frac{\me^{- u}}{-\mi u/t -\omega_{\mS}+B
   \me^{3\mi(1-\nu)\pi/2} (u/t)^{\nu-1}}\Bigg).\label{54}
\end{align}
Due to the complexity of evaluating the integral above, we employ the large-time asymptotic analysis used in Refs.~\cite{garmon2013amplification, taira2024markovianity}. For large $t$, the dominant term of the denominator is $(u/t)^{\nu-1}$ for $0<\nu<1$. Since we are interested in the long-time dynamics, we neglect the other terms $\mi u/t$ and $\omega_{\mS}$ as follows:
\begin{align}
    \int _{C_\mathrm{a}+C_\mathrm{b}}\dd{z} F(z)
    &\simeq \frac{\mi}{t } \int_{0}^{\infty}\dd{u} \Bigg(\frac{\me^{- u}}{B
   \me^{-\mi(1-\nu)\pi/2} (u/t)^{\nu-1}}-\frac{\me^{- u}}{B
   \me^{3\mi(1-\nu)\pi/2} (u/t)^{\nu-1}}\Bigg)\label{55}\\
   &=\frac{\mi t^{\nu-2}}{B}\qty[ \me^{\mi(1-\nu)\pi/2}-\me^{-3\mi(1-\nu)\pi/2}] \int_{0}^{\infty}\dd{u} \me^{- u}u^{1-\nu}\label{56}\\
   &=\frac{\mi t^{\nu-2}}{B}\qty[ \me^{\mi(1-\nu)\pi/2}-\me^{-3\mi(1-\nu)\pi/2} ]\Gamma(2-\nu),\label{57}
  \end{align}
where we have used the formula of the gamma function \cite{gradshteyn2014table}.
Therefore, we obtain the integral of the asymptotic expansion in terms of the leading order of $1/t$ along the paths $C_a+C_b$, which is the same as Eq.~(37) in the main text.

\section{D. Time-local master equation and relation to the GKSL equation}

In this section, we clarify the relation between the survival amplitude analyzed in the main text and the GKSL equation. This relation is standard for spontaneous-emission models within the rotating-wave approximation \cite{breuer2002theory,garraway1997nonperturbative}.

We consider the same Hamiltonian as in Eq.~(8) of the main text. In addition to the single-excitation states
\begin{gather}
    \ket{\mS}:=\ket{1}_{\rm S}\otimes\ket{\mathrm{vac}}_{\rm E},
    \qquad
    \ket{\vec{k}}:=\ket{0}_{\rm S}\otimes\ket{\vec{k}}_{\rm E},
\end{gather}
we also introduce the zero-excitation state
\begin{gather}
    \ket{G}:=\ket{0}_{\rm S}\otimes\ket{\mathrm{vac}}_{\rm E}.
\end{gather}
Because of the rotating-wave form of the interaction Hamiltonian, the state \(\ket{G}\) is invariant under the time evolution.

 In this setup, the time-evolved total state can be written as
\begin{gather}
    \ket{\Psi(t)}
    =
    c_{\rm S}(t)\ket{\mS}
    +
    \int_{\mathbb{R}^D}\dd{\vec{k}}\,
    c_{\vec{k}}(t)\ket{\vec{k}}
    +
    c_0\ket{G}.
\end{gather}

Tracing out the environment, we obtain the reduced density matrix of the two-level system:
\begin{align}
    \rho_{\rm S}(t)
    &=
    \Tr_{\rm E}\ketbra{\Psi(t)}
    \notag\\
    &=
    \abs{c_{\rm S}(t)}^2\ketbra{1}
    +
    c_0^*c_{\rm S}(t)\ketbra*{1}{0}
    +
    c_0c_{\rm S}^*(t)\ketbra*{0}{1}
    +
    \qty(1-\abs{c_{\rm S}(t)}^2)\ketbra{0}.
\end{align}
Equivalently, in the basis \(\{\ket{1},\ket{0}\}\), this reads
\begin{gather}
    \rho_{\rm S}(t)
    =
    \mqty(
    \abs{c_{\rm S}(t)}^2
    &
    c_0^*c_{\rm S}(t)
    \\
    c_0c_{\rm S}^*(t)
    &
    1-\abs{c_{\rm S}(t)}^2
    ).
    \label{eq:rho_matrix_gksl_form}
\end{gather}
Differentiating Eq.~\eqref{eq:rho_matrix_gksl_form} with respect to time, we obtain
\begin{gather}
    \dv{}{t}\rho_{\rm S}(t)
    =
    \mqty(
    \dv{}{t}\abs{c_{\rm S}(t)}^2
    &
    c_0^*\dot{c}_{\rm S}(t)
    \\
    c_0\dot{c}_{\rm S}^*(t)
    &
    -\dv{}{t}\abs{c_{\rm S}(t)}^2
    ).
    \label{eq:rho_dot_matrix_gksl_form}
\end{gather}
Whenever \(c_{\rm S}(t)\neq0\), Eq.~\eqref{eq:rho_dot_matrix_gksl_form} can be rewritten as
\begin{align}
    \dv{}{t}\rho_{\rm S}(t)
    =
    -\frac{\mi}{2}S(t)
    \qty[\sigma_+\sigma_-,\rho_{\rm S}(t)]
    +
    \gamma(t)
    \qty(
    \sigma_-\rho_{\rm S}(t)\sigma_+
    -
    \frac{1}{2}
    \qty{\sigma_+\sigma_-,\rho_{\rm S}(t)}
    ),
    \label{eq:time_local_master}
\end{align}
where
\begin{gather}
    S(t):=-2\Im\frac{\dot{c}_{\rm S}(t)}{c_{\rm S}(t)},
    \qquad
    \gamma(t):=-2\Re\frac{\dot{c}_{\rm S}(t)}{c_{\rm S}(t)} .
    \label{eq:time_dependent_coefficients}
\end{gather}
Thus the exact reduced dynamics has the same operator structure as the energy-relaxation GKSL equation, but with time-dependent coefficients.

If the survival amplitude is a single exponential,
\begin{gather}
    c_{\rm S}(t)
    =
    c_{\rm S}(0)
    \exp[-(\gamma/2+\mi S/2)t],
\end{gather}
then \(\gamma(t)=\gamma\) and \(S(t)=S\) are constants. In this case, Eq.~\eqref{eq:time_local_master} reduces to the time-independent GKSL equation
\begin{align}
    \dv{}{t}\rho_{\rm S}(t)
    =
    -\frac{\mi}{2}S
    \qty[\sigma_+\sigma_-,\rho_{\rm S}(t)]
    +
    \gamma
    \qty(
    \sigma_-\rho_{\rm S}(t)\sigma_+
    -
    \frac{1}{2}
    \qty{\sigma_+\sigma_-,\rho_{\rm S}(t)}
    ),
    \label{eq:time_independent_gksl}
\end{align}
which generates a one-parameter semigroup. Conversely, if \(c_{\rm S}(t)\) is not a single exponential, the coefficients \(S(t)\) and \(\gamma(t)\) are generally time dependent, and the reduced dynamics is not described by a time-independent GKSL equation.

Therefore the fractional power-law behavior derived in the main text corresponds to time-dependent coefficients in the exact time-local master equation. This shows explicitly that the fractional power-law decay cannot be reproduced by a time-independent GKSL equation.

For the initially excited state $c_{\rm S}(0)=1$ and $c_0=0$, the time-independent GKSL equation gives only exponential decay. Indeed, substituting
\begin{gather}
    \rho_{\rm S}(t)
    =
    \mqty(
    \abs{c_{\rm S}(t)}^2 & 0\\
    0 & 1-\abs{c_{\rm S}(t)}^2
    )
\end{gather}
into Eq.~\eqref{eq:time_independent_gksl}, we obtain
\begin{gather}
    \dv{}{t}\abs{c_{\rm S}(t)}^2
    =
    -\gamma \abs{c_{\rm S}(t)}^2,
    \qquad
    \abs{c_{\rm S}(0)}^2=1 .
\end{gather}
Hence
\begin{gather}
    \abs{c_{\rm S}(t)}^2=\me^{-\gamma t}.
\end{gather}
Therefore, a time-independent GKSL equation only reproduces the purely exponential decay.

\section{E. COUNTER-ROTATING TERMS AT ORDER \(g^2\)}

In the main text, we have used the interaction Hamiltonian within the rotating-wave approximation. Here we show that the counter-rotating terms do not contribute to the survival amplitude up to order \(g^2\) for the initial state considered in this work.

Let us consider the interaction Hamiltonian before applying the RWA:
\begin{align}
    H_{\mathrm{int}}^{\mathrm{full}}
    =
    g\int_{\mathbb{R}^D}\dd{\vec{k}}\,
    \left[
        \sigma_+ b(\vec{k})
        + \sigma_- b^\dagger(\vec{k})
        + \sigma_+ b^\dagger(\vec{k})
        + \sigma_- b(\vec{k})
    \right].
\end{align}
The last two terms are the counter-rotating terms. We transform this Hamiltonian into the interaction picture with respect to
\begin{align}
    H_0
    =
    \omega_{\mathrm{S}}\sigma_+\sigma_- \otimes I_{\rm E}
    +
    I_{\mS}\otimes \omega_0
    \int_{\mathbb{R}^D}\dd{\vec{k}}\,
    |\vec{k}|^n
    b^\dagger(\vec{k})b(\vec{k}) .
\end{align}
Then, we obtain
\begin{align}
    H_{\mathrm{int}}^{\mathrm{full},(I)}(t)
    &:=
    \me^{\mi H_0t}
    H_{\mathrm{int}}^{\mathrm{full}}
    \me^{-\mi H_0t}
    \nonumber\\
    &=
    g\int_{\mathbb{R}^D}\dd{\vec{k}}\,
    \Big[
        \me^{\mi(\omega_{\mathrm{S}}-\omega_0|\vec{k}|^n)t}
        \sigma_+ b(\vec{k})
        +
        \me^{-\mi(\omega_{\mathrm{S}}-\omega_0|\vec{k}|^n)t}
        \sigma_- b^\dagger(\vec{k})
        \nonumber\\
        &\hspace{3.2cm}
        +
        \me^{\mi(\omega_{\mathrm{S}}+\omega_0|\vec{k}|^n)t}
        \sigma_+ b^\dagger(\vec{k})
        +
        \me^{-\mi(\omega_{\mathrm{S}}+\omega_0|\vec{k}|^n)t}
        \sigma_- b(\vec{k})
    \Big].
\end{align}
For the initial state
\begin{align}
    \ket{\mS}
    =
    \ket{1}_{\mathrm{S}}\otimes\ket{\mathrm{vac}}_{\mathrm{E}},
\end{align}
the first-order contribution of the Dyson series to the survival amplitude vanishes. The second-order contribution is determined by
\begin{align}
    \eta_{\mathrm{full}}(t_1-t_2)
    :=
    \bra{\mS}
    H_{\mathrm{int}}^{\mathrm{full},(I)}(t_1)
    H_{\mathrm{int}}^{\mathrm{full},(I)}(t_2)
    \ket{\mS}.
\end{align}
When \(H_{\mathrm{int}}^{\mathrm{full},(I)}(t_2)\) acts on \(\ket{\mS}\), only the term
\(\sigma_- b^\dagger(\vec{k})\) gives a nonzero contribution:
\begin{align}
    H_{\mathrm{int}}^{\mathrm{full},(I)}(t_2)\ket{\mS}
    =
    g\int_{\mathbb{R}^D}\dd{\vec{k}}\,
    \me^{-\mi(\omega_{\mathrm{S}}-\omega_0|\vec{k}|^n)t_2}
    \ket{0}_{\mathrm{S}}\otimes\ket{\vec{k}}_{\mathrm{E}} .
\end{align}
Indeed, the terms \(\sigma_+ b(\vec{k})\) and \(\sigma_+ b^\dagger(\vec{k})\) vanish because
\(\sigma_+\ket{1}_{\mathrm{S}}=0\), while the term \(\sigma_- b(\vec{k})\) vanishes because
\(b(\vec{k})\ket{\mathrm{vac}}_{\mathrm{E}}=0\). To return to \(\ket{\mS}\) at time \(t_1\), only the term
\(\sigma_+ b(\vec{k})\) can contribute. Therefore,
\begin{align}
    \eta_{\mathrm{full}}(t_1-t_2)
    &=
    g^2
    \int_{\mathbb{R}^D}\dd{\vec{k}}\,
    \me^{\mi(\omega_{\mathrm{S}}-\omega_0|\vec{k}|^n)(t_1-t_2)}
    \nonumber\\
    &=
    \eta(t_1-t_2),
\end{align}
where \(\eta(t_1-t_2)\) is the autocorrelation function obtained in Eq.~(S17). Hence the counter-rotating terms do not affect the \(O(g^2)\) contribution to the survival amplitude.

Thus, the leading short-time fractional power-law term is unchanged at order \(g^2\). The long-time analysis, however, is based on the rotating-wave approximation Hamiltonian, and a full beyond rotating-wave approximation treatment of the long-time pole structure is beyond the scope of the present work.
\nocite{*}
\bibliographystyle{apsrev4-2}
\bibliography{Nakabayashi_Supplementary_2.bib}

\begin{thebibliography}{36}%
\makeatletter
\providecommand \@ifxundefined [1]{%
 \@ifx{#1\undefined}
}%
\providecommand \@ifnum [1]{%
 \ifnum #1\expandafter \@firstoftwo
 \else \expandafter \@secondoftwo
 \fi
}%
\providecommand \@ifx [1]{%
 \ifx #1\expandafter \@firstoftwo
 \else \expandafter \@secondoftwo
 \fi
}%
\providecommand \natexlab [1]{#1}%
\providecommand \enquote  [1]{``#1''}%
\providecommand \bibnamefont  [1]{#1}%
\providecommand \bibfnamefont [1]{#1}%
\providecommand \citenamefont [1]{#1}%
\providecommand \href@noop [0]{\@secondoftwo}%
\providecommand \href [0]{\begingroup \@sanitize@url \@href}%
\providecommand \@href[1]{\@@startlink{#1}\@@href}%
\providecommand \@@href[1]{\endgroup#1\@@endlink}%
\providecommand \@sanitize@url [0]{\catcode `\\12\catcode `\$12\catcode
  `\&12\catcode `\#12\catcode `\^12\catcode `\_12\catcode `\%12\relax}%
\providecommand \@@startlink[1]{}%
\providecommand \@@endlink[0]{}%
\providecommand \url  [0]{\begingroup\@sanitize@url \@url }%
\providecommand \@url [1]{\endgroup\@href {#1}{\urlprefix }}%
\providecommand \urlprefix  [0]{URL }%
\providecommand \Eprint [0]{\href }%
\providecommand \doibase [0]{https://doi.org/}%
\providecommand \selectlanguage [0]{\@gobble}%
\providecommand \bibinfo  [0]{\@secondoftwo}%
\providecommand \bibfield  [0]{\@secondoftwo}%
\providecommand \translation [1]{[#1]}%
\providecommand \BibitemOpen [0]{}%
\providecommand \bibitemStop [0]{}%
\providecommand \bibitemNoStop [0]{.\EOS\space}%
\providecommand \EOS [0]{\spacefactor3000\relax}%
\providecommand \BibitemShut  [1]{\csname bibitem#1\endcsname}%
\let\auto@bib@innerbib\@empty
\bibitem [{\citenamefont {Gamow}(1928)}]{gamow}%
  \BibitemOpen
  \bibfield  {author} {\bibinfo {author} {\bibfnamefont {G.}~\bibnamefont
  {Gamow}},\ }\bibfield  {title} {\bibinfo {title} {{Zur Quantentheorie des
  Atomkernes}},\ }\href@noop {} {\bibfield  {journal} {\bibinfo  {journal} {Z.
  Physik}\ }\textbf {\bibinfo {volume} {51}},\ \bibinfo {pages} {204} (\bibinfo
  {year} {1928})}\BibitemShut {NoStop}%
\bibitem [{\citenamefont {Khalfin}(1958)}]{khalfin1958contribution}%
  \BibitemOpen
  \bibfield  {author} {\bibinfo {author} {\bibfnamefont {L.}~\bibnamefont
  {Khalfin}},\ }\bibfield  {title} {\bibinfo {title} {{Contribution to the
  decay theory of a quasi-stationary state}},\ }\href@noop {} {\bibfield
  {journal} {\bibinfo  {journal} {Soviet Physics JETP}\ }\textbf {\bibinfo
  {volume} {6}},\ \bibinfo {pages} {1053} (\bibinfo {year} {1958})}\BibitemShut
  {NoStop}%
\bibitem [{\citenamefont {Fonda}\ \emph {et~al.}(1978)\citenamefont {Fonda},
  \citenamefont {Ghirardi},\ and\ \citenamefont {Rimini}}]{fonda1978decay}%
  \BibitemOpen
  \bibfield  {author} {\bibinfo {author} {\bibfnamefont {L.}~\bibnamefont
  {Fonda}}, \bibinfo {author} {\bibfnamefont {G.}~\bibnamefont {Ghirardi}},\
  and\ \bibinfo {author} {\bibfnamefont {A.}~\bibnamefont {Rimini}},\
  }\bibfield  {title} {\bibinfo {title} {{Decay theory of unstable quantum
  systems}},\ }\href@noop {} {\bibfield  {journal} {\bibinfo  {journal}
  {Reports on Progress in Physics}\ }\textbf {\bibinfo {volume} {41}},\
  \bibinfo {pages} {587} (\bibinfo {year} {1978})}\BibitemShut {NoStop}%
\bibitem [{\citenamefont {Peshkin}\ \emph {et~al.}(2014)\citenamefont
  {Peshkin}, \citenamefont {Volya},\ and\ \citenamefont
  {Zelevinsky}}]{peshkin2014non}%
  \BibitemOpen
  \bibfield  {author} {\bibinfo {author} {\bibfnamefont {M.}~\bibnamefont
  {Peshkin}}, \bibinfo {author} {\bibfnamefont {A.}~\bibnamefont {Volya}},\
  and\ \bibinfo {author} {\bibfnamefont {V.}~\bibnamefont {Zelevinsky}},\
  }\bibfield  {title} {\bibinfo {title} {{Non-exponential and oscillatory
  decays in quantum mechanics}},\ }\href@noop {} {\bibfield  {journal}
  {\bibinfo  {journal} {Europhysics Letters}\ }\textbf {\bibinfo {volume}
  {107}},\ \bibinfo {pages} {40001} (\bibinfo {year} {2014})}\BibitemShut
  {NoStop}%
\bibitem [{\citenamefont {Chiu}\ \emph {et~al.}(1977)\citenamefont {Chiu},
  \citenamefont {Sudarshan},\ and\ \citenamefont {Misra}}]{chiu1977time}%
  \BibitemOpen
  \bibfield  {author} {\bibinfo {author} {\bibfnamefont {C.}~\bibnamefont
  {Chiu}}, \bibinfo {author} {\bibfnamefont {E.}~\bibnamefont {Sudarshan}},\
  and\ \bibinfo {author} {\bibfnamefont {B.}~\bibnamefont {Misra}},\ }\bibfield
   {title} {\bibinfo {title} {{Time evolution of unstable quantum states and a
  resolution of Zeno's paradox}},\ }\href@noop {} {\bibfield  {journal}
  {\bibinfo  {journal} {Physical Review D}\ }\textbf {\bibinfo {volume} {16}},\
  \bibinfo {pages} {520} (\bibinfo {year} {1977})}\BibitemShut {NoStop}%
\bibitem [{\citenamefont {Wilkinson}\ \emph {et~al.}(1997)\citenamefont
  {Wilkinson}, \citenamefont {Bharucha}, \citenamefont {Fischer}, \citenamefont
  {Madison}, \citenamefont {Morrow}, \citenamefont {Niu}, \citenamefont
  {Sundaram},\ and\ \citenamefont {Raizen}}]{wilkinson1997experimental}%
  \BibitemOpen
  \bibfield  {author} {\bibinfo {author} {\bibfnamefont {S.~R.}\ \bibnamefont
  {Wilkinson}}, \bibinfo {author} {\bibfnamefont {C.~F.}\ \bibnamefont
  {Bharucha}}, \bibinfo {author} {\bibfnamefont {M.~C.}\ \bibnamefont
  {Fischer}}, \bibinfo {author} {\bibfnamefont {K.~W.}\ \bibnamefont
  {Madison}}, \bibinfo {author} {\bibfnamefont {P.~R.}\ \bibnamefont {Morrow}},
  \bibinfo {author} {\bibfnamefont {Q.}~\bibnamefont {Niu}}, \bibinfo {author}
  {\bibfnamefont {B.}~\bibnamefont {Sundaram}},\ and\ \bibinfo {author}
  {\bibfnamefont {M.~G.}\ \bibnamefont {Raizen}},\ }\bibfield  {title}
  {\bibinfo {title} {{Experimental evidence for non-exponential decay in
  quantum tunnelling}},\ }\href@noop {} {\bibfield  {journal} {\bibinfo
  {journal} {Nature}\ }\textbf {\bibinfo {volume} {387}},\ \bibinfo {pages}
  {575} (\bibinfo {year} {1997})}\BibitemShut {NoStop}%
\bibitem [{\citenamefont {Rothe}\ \emph {et~al.}(2006)\citenamefont {Rothe},
  \citenamefont {Hintschich},\ and\ \citenamefont
  {Monkman}}]{rothe2006violation}%
  \BibitemOpen
  \bibfield  {author} {\bibinfo {author} {\bibfnamefont {C.}~\bibnamefont
  {Rothe}}, \bibinfo {author} {\bibfnamefont {S.}~\bibnamefont {Hintschich}},\
  and\ \bibinfo {author} {\bibfnamefont {A.}~\bibnamefont {Monkman}},\
  }\bibfield  {title} {\bibinfo {title} {{Violation of the exponential-decay
  law at long times}},\ }\href@noop {} {\bibfield  {journal} {\bibinfo
  {journal} {Physical Review Letters}\ }\textbf {\bibinfo {volume} {96}},\
  \bibinfo {pages} {163601} (\bibinfo {year} {2006})}\BibitemShut {NoStop}%
\bibitem [{\citenamefont {Crespi}\ \emph {et~al.}(2019)\citenamefont {Crespi},
  \citenamefont {Pepe}, \citenamefont {Facchi}, \citenamefont {Sciarrino},
  \citenamefont {Mataloni}, \citenamefont {Nakazato}, \citenamefont
  {Pascazio},\ and\ \citenamefont {Osellame}}]{crespi2019experimental}%
  \BibitemOpen
  \bibfield  {author} {\bibinfo {author} {\bibfnamefont {A.}~\bibnamefont
  {Crespi}}, \bibinfo {author} {\bibfnamefont {F.~V.}\ \bibnamefont {Pepe}},
  \bibinfo {author} {\bibfnamefont {P.}~\bibnamefont {Facchi}}, \bibinfo
  {author} {\bibfnamefont {F.}~\bibnamefont {Sciarrino}}, \bibinfo {author}
  {\bibfnamefont {P.}~\bibnamefont {Mataloni}}, \bibinfo {author}
  {\bibfnamefont {H.}~\bibnamefont {Nakazato}}, \bibinfo {author}
  {\bibfnamefont {S.}~\bibnamefont {Pascazio}},\ and\ \bibinfo {author}
  {\bibfnamefont {R.}~\bibnamefont {Osellame}},\ }\bibfield  {title} {\bibinfo
  {title} {{Experimental investigation of quantum decay at short, intermediate,
  and long times via integrated photonics}},\ }\href@noop {} {\bibfield
  {journal} {\bibinfo  {journal} {Physical Review Letters}\ }\textbf {\bibinfo
  {volume} {122}},\ \bibinfo {pages} {130401} (\bibinfo {year}
  {2019})}\BibitemShut {NoStop}%
\bibitem [{\citenamefont {Misra}\ and\ \citenamefont
  {Sudarshan}(1977)}]{misra1977zeno}%
  \BibitemOpen
  \bibfield  {author} {\bibinfo {author} {\bibfnamefont {B.}~\bibnamefont
  {Misra}}\ and\ \bibinfo {author} {\bibfnamefont {E.~G.}\ \bibnamefont
  {Sudarshan}},\ }\bibfield  {title} {\bibinfo {title} {{The Zeno's paradox in
  quantum theory}},\ }\href@noop {} {\bibfield  {journal} {\bibinfo  {journal}
  {Journal of Mathematical Physics}\ }\textbf {\bibinfo {volume} {18}},\
  \bibinfo {pages} {756} (\bibinfo {year} {1977})}\BibitemShut {NoStop}%
\bibitem [{\citenamefont {Itano}\ \emph {et~al.}(1990)\citenamefont {Itano},
  \citenamefont {Heinzen}, \citenamefont {Bollinger},\ and\ \citenamefont
  {Wineland}}]{itano1990quantum}%
  \BibitemOpen
  \bibfield  {author} {\bibinfo {author} {\bibfnamefont {W.~M.}\ \bibnamefont
  {Itano}}, \bibinfo {author} {\bibfnamefont {D.~J.}\ \bibnamefont {Heinzen}},
  \bibinfo {author} {\bibfnamefont {J.~J.}\ \bibnamefont {Bollinger}},\ and\
  \bibinfo {author} {\bibfnamefont {D.~J.}\ \bibnamefont {Wineland}},\
  }\bibfield  {title} {\bibinfo {title} {{Quantum Zeno effect}},\ }\href@noop
  {} {\bibfield  {journal} {\bibinfo  {journal} {Physical Review A}\ }\textbf
  {\bibinfo {volume} {41}},\ \bibinfo {pages} {2295} (\bibinfo {year}
  {1990})}\BibitemShut {NoStop}%
\bibitem [{\citenamefont {Kofman}\ \emph {et~al.}(1994)\citenamefont {Kofman},
  \citenamefont {Kurizki},\ and\ \citenamefont
  {Sherman}}]{kofman1994spontaneous}%
  \BibitemOpen
  \bibfield  {author} {\bibinfo {author} {\bibfnamefont {A.}~\bibnamefont
  {Kofman}}, \bibinfo {author} {\bibfnamefont {G.}~\bibnamefont {Kurizki}},\
  and\ \bibinfo {author} {\bibfnamefont {B.}~\bibnamefont {Sherman}},\
  }\bibfield  {title} {\bibinfo {title} {{Spontaneous and induced atomic decay
  in photonic band structures}},\ }\href@noop {} {\bibfield  {journal}
  {\bibinfo  {journal} {Journal of Modern Optics}\ }\textbf {\bibinfo {volume}
  {41}},\ \bibinfo {pages} {353} (\bibinfo {year} {1994})}\BibitemShut
  {NoStop}%
\bibitem [{\citenamefont {Gorini}\ \emph {et~al.}(1976)\citenamefont {Gorini},
  \citenamefont {Kossakowski},\ and\ \citenamefont
  {Sudarshan}}]{gorini1976completely}%
  \BibitemOpen
  \bibfield  {author} {\bibinfo {author} {\bibfnamefont {V.}~\bibnamefont
  {Gorini}}, \bibinfo {author} {\bibfnamefont {A.}~\bibnamefont
  {Kossakowski}},\ and\ \bibinfo {author} {\bibfnamefont {E.~C.~G.}\
  \bibnamefont {Sudarshan}},\ }\bibfield  {title} {\bibinfo {title}
  {{Completely positive dynamical semigroups of N-level systems}},\ }\href@noop
  {} {\bibfield  {journal} {\bibinfo  {journal} {Journal of Mathematical
  Physics}\ }\textbf {\bibinfo {volume} {17}},\ \bibinfo {pages} {821}
  (\bibinfo {year} {1976})}\BibitemShut {NoStop}%
\bibitem [{\citenamefont {Lindblad}(1976)}]{lindblad1976generators}%
  \BibitemOpen
  \bibfield  {author} {\bibinfo {author} {\bibfnamefont {G.}~\bibnamefont
  {Lindblad}},\ }\bibfield  {title} {\bibinfo {title} {{On the generators of
  quantum dynamical semigroups}},\ }\href@noop {} {\bibfield  {journal}
  {\bibinfo  {journal} {Communications in Mathematical Physics}\ }\textbf
  {\bibinfo {volume} {48}},\ \bibinfo {pages} {119} (\bibinfo {year}
  {1976})}\BibitemShut {NoStop}%
\bibitem [{\citenamefont {Taira}\ \emph {et~al.}(2024)\citenamefont {Taira},
  \citenamefont {Hatano},\ and\ \citenamefont
  {Nishino}}]{taira2024markovianity}%
  \BibitemOpen
  \bibfield  {author} {\bibinfo {author} {\bibfnamefont {T.}~\bibnamefont
  {Taira}}, \bibinfo {author} {\bibfnamefont {N.}~\bibnamefont {Hatano}},\ and\
  \bibinfo {author} {\bibfnamefont {A.}~\bibnamefont {Nishino}},\ }\bibfield
  {title} {\bibinfo {title} {{Markovianity and non-Markovianity of Particle
  Bath with Dirac Dispersion Relation}},\ }\href@noop {} {\bibfield  {journal}
  {\bibinfo  {journal} {arXiv preprint arXiv:2406.17436}\ } (\bibinfo {year}
  {2024})}\BibitemShut {NoStop}%
\bibitem [{\citenamefont {Facchi}\ and\ \citenamefont
  {Pascazio}(2008)}]{facchi2008quantum}%
  \BibitemOpen
  \bibfield  {author} {\bibinfo {author} {\bibfnamefont {P.}~\bibnamefont
  {Facchi}}\ and\ \bibinfo {author} {\bibfnamefont {S.}~\bibnamefont
  {Pascazio}},\ }\bibfield  {title} {\bibinfo {title} {{Quantum Zeno dynamics:
  mathematical and physical aspects}},\ }\href@noop {} {\bibfield  {journal}
  {\bibinfo  {journal} {Journal of Physics A: Mathematical and Theoretical}\
  }\textbf {\bibinfo {volume} {41}},\ \bibinfo {pages} {493001} (\bibinfo
  {year} {2008})}\BibitemShut {NoStop}%
\bibitem [{\citenamefont {Nishino}\ and\ \citenamefont
  {Hatano}(2024)}]{nishino2024exact}%
  \BibitemOpen
  \bibfield  {author} {\bibinfo {author} {\bibfnamefont {A.}~\bibnamefont
  {Nishino}}\ and\ \bibinfo {author} {\bibfnamefont {N.}~\bibnamefont
  {Hatano}},\ }\bibfield  {title} {\bibinfo {title} {{Exact time-evolving
  scattering states in open quantum-dot systems with an interaction: discovery
  of time-evolving resonant states}},\ }\href@noop {} {\bibfield  {journal}
  {\bibinfo  {journal} {Journal of Physics A: Mathematical and Theoretical}\
  }\textbf {\bibinfo {volume} {57}},\ \bibinfo {pages} {245302} (\bibinfo
  {year} {2024})}\BibitemShut {NoStop}%
\bibitem [{\citenamefont {Feshbach}(1958)}]{feshbach1958unified}%
  \BibitemOpen
  \bibfield  {author} {\bibinfo {author} {\bibfnamefont {H.}~\bibnamefont
  {Feshbach}},\ }\bibfield  {title} {\bibinfo {title} {{Unified theory of
  nuclear reactions}},\ }\href@noop {} {\bibfield  {journal} {\bibinfo
  {journal} {Annals of Physics}\ }\textbf {\bibinfo {volume} {5}},\ \bibinfo
  {pages} {357} (\bibinfo {year} {1958})}\BibitemShut {NoStop}%
\bibitem [{\citenamefont {Hatano}\ and\ \citenamefont
  {Ordonez}(2014)}]{hatano2014time}%
  \BibitemOpen
  \bibfield  {author} {\bibinfo {author} {\bibfnamefont {N.}~\bibnamefont
  {Hatano}}\ and\ \bibinfo {author} {\bibfnamefont {G.}~\bibnamefont
  {Ordonez}},\ }\bibfield  {title} {\bibinfo {title} {{Time-reversal symmetric
  resolution of unity without background integrals in open quantum systems}},\
  }\href@noop {} {\bibfield  {journal} {\bibinfo  {journal} {Journal of
  Mathematical Physics}\ }\textbf {\bibinfo {volume} {55}} (\bibinfo {year}
  {2014})}\BibitemShut {NoStop}%
\bibitem [{\citenamefont {Dantas}\ \emph {et~al.}(2020)\citenamefont {Dantas},
  \citenamefont {Pe{\~n}a-Benitez}, \citenamefont {Roy},\ and\ \citenamefont
  {Sur{\'o}wka}}]{dantas2020non}%
  \BibitemOpen
  \bibfield  {author} {\bibinfo {author} {\bibfnamefont {R.~M.}\ \bibnamefont
  {Dantas}}, \bibinfo {author} {\bibfnamefont {F.}~\bibnamefont
  {Pe{\~n}a-Benitez}}, \bibinfo {author} {\bibfnamefont {B.}~\bibnamefont
  {Roy}},\ and\ \bibinfo {author} {\bibfnamefont {P.}~\bibnamefont
  {Sur{\'o}wka}},\ }\bibfield  {title} {\bibinfo {title} {{Non-Abelian
  anomalies in multi-Weyl semimetals}},\ }\href@noop {} {\bibfield  {journal}
  {\bibinfo  {journal} {Physical Review Research}\ }\textbf {\bibinfo {volume}
  {2}},\ \bibinfo {pages} {013007} (\bibinfo {year} {2020})}\BibitemShut
  {NoStop}%
\bibitem [{\citenamefont {Schmeltzer}\ and\ \citenamefont
  {Saxena}(2023)}]{schmeltzer2023optical}%
  \BibitemOpen
  \bibfield  {author} {\bibinfo {author} {\bibfnamefont {D.}~\bibnamefont
  {Schmeltzer}}\ and\ \bibinfo {author} {\bibfnamefont {A.}~\bibnamefont
  {Saxena}},\ }\bibfield  {title} {\bibinfo {title} {{Optical conductivity of
  multi-Weyl node semimetals due to electromagnetic and axial pseudogauge
  fields}},\ }\href@noop {} {\bibfield  {journal} {\bibinfo  {journal} {Annals
  of Physics}\ }\textbf {\bibinfo {volume} {455}},\ \bibinfo {pages} {169380}
  (\bibinfo {year} {2023})}\BibitemShut {NoStop}%
\bibitem [{\citenamefont {Burr}\ and\ \citenamefont
  {Reano}(2015)}]{burr2015zero}%
  \BibitemOpen
  \bibfield  {author} {\bibinfo {author} {\bibfnamefont {J.~R.}\ \bibnamefont
  {Burr}}\ and\ \bibinfo {author} {\bibfnamefont {R.~M.}\ \bibnamefont
  {Reano}},\ }\bibfield  {title} {\bibinfo {title} {{Zero-coupling-gap
  degenerate band edge resonators in silicon photonics}},\ }\href@noop {}
  {\bibfield  {journal} {\bibinfo  {journal} {Optics Express}\ }\textbf
  {\bibinfo {volume} {23}},\ \bibinfo {pages} {30933} (\bibinfo {year}
  {2015})}\BibitemShut {NoStop}%
\bibitem [{\citenamefont {Wood}\ \emph {et~al.}(2015)\citenamefont {Wood},
  \citenamefont {Burr},\ and\ \citenamefont {Reano}}]{wood2015degenerate}%
  \BibitemOpen
  \bibfield  {author} {\bibinfo {author} {\bibfnamefont {M.~G.}\ \bibnamefont
  {Wood}}, \bibinfo {author} {\bibfnamefont {J.~R.}\ \bibnamefont {Burr}},\
  and\ \bibinfo {author} {\bibfnamefont {R.~M.}\ \bibnamefont {Reano}},\
  }\bibfield  {title} {\bibinfo {title} {{Degenerate band edge resonances in
  periodic silicon ridge waveguides}},\ }\href@noop {} {\bibfield  {journal}
  {\bibinfo  {journal} {Optics Letters}\ }\textbf {\bibinfo {volume} {40}},\
  \bibinfo {pages} {2493} (\bibinfo {year} {2015})}\BibitemShut {NoStop}%
\bibitem [{\citenamefont {Yazdi}\ \emph {et~al.}(2022)\citenamefont {Yazdi},
  \citenamefont {Oshmarin}, \citenamefont {Mealy}, \citenamefont {Almutawa},
  \citenamefont {Nikzamir},\ and\ \citenamefont {Capolino}}]{yazdi2022sixth}%
  \BibitemOpen
  \bibfield  {author} {\bibinfo {author} {\bibfnamefont {F.}~\bibnamefont
  {Yazdi}}, \bibinfo {author} {\bibfnamefont {D.}~\bibnamefont {Oshmarin}},
  \bibinfo {author} {\bibfnamefont {T.}~\bibnamefont {Mealy}}, \bibinfo
  {author} {\bibfnamefont {A.~T.}\ \bibnamefont {Almutawa}}, \bibinfo {author}
  {\bibfnamefont {A.}~\bibnamefont {Nikzamir}},\ and\ \bibinfo {author}
  {\bibfnamefont {F.}~\bibnamefont {Capolino}},\ }\bibfield  {title} {\bibinfo
  {title} {Sixth-order degenerate band edge in coupled microstrip waveguides},\
  }\href@noop {} {\bibfield  {journal} {\bibinfo  {journal} {Physical Review
  Applied}\ }\textbf {\bibinfo {volume} {17}},\ \bibinfo {pages} {064049}
  (\bibinfo {year} {2022})}\BibitemShut {NoStop}%
\bibitem [{\citenamefont {Breuer}\ and\ \citenamefont
  {Petruccione}(2002)}]{breuer2002theory}%
  \BibitemOpen
  \bibfield  {author} {\bibinfo {author} {\bibfnamefont {H.-P.}\ \bibnamefont
  {Breuer}}\ and\ \bibinfo {author} {\bibfnamefont {F.}~\bibnamefont
  {Petruccione}},\ }\href@noop {} {\emph {\bibinfo {title} {{The theory of open
  quantum systems}}}}\ (\bibinfo  {publisher} {OUP Oxford},\ \bibinfo {year}
  {2002})\BibitemShut {NoStop}%
\bibitem [{\citenamefont {Garraway}(1997)}]{garraway1997nonperturbative}%
  \BibitemOpen
  \bibfield  {author} {\bibinfo {author} {\bibfnamefont {B.~M.}\ \bibnamefont
  {Garraway}},\ }\bibfield  {title} {\bibinfo {title} {Nonperturbative decay of
  an atomic system in a cavity},\ }\href@noop {} {\bibfield  {journal}
  {\bibinfo  {journal} {Physical Review A}\ }\textbf {\bibinfo {volume} {55}},\
  \bibinfo {pages} {2290} (\bibinfo {year} {1997})}\BibitemShut {NoStop}%
\bibitem [{sup()}]{supplemental_material}%
  \BibitemOpen
  \href@noop {} {}\bibinfo {note} {See Supplemental Material at [URL will be
  inserted by publisher] for details of backgrounds, proofs, and extended
  discussions, which includes Ref.~\cite{gradshteyn2014table}}\BibitemShut
  {NoStop}%
\bibitem [{\citenamefont {Gradshteyn}\ and\ \citenamefont
  {Ryzhik}(2014)}]{gradshteyn2014table}%
  \BibitemOpen
  \bibfield  {author} {\bibinfo {author} {\bibfnamefont {I.~S.}\ \bibnamefont
  {Gradshteyn}}\ and\ \bibinfo {author} {\bibfnamefont {I.~M.}\ \bibnamefont
  {Ryzhik}},\ }\href@noop {} {\emph {\bibinfo {title} {{Table of integrals,
  series, and products}}}}\ (\bibinfo  {publisher} {Academic press},\ \bibinfo
  {year} {2014})\BibitemShut {NoStop}%
\bibitem [{\citenamefont {Petrosky}\ and\ \citenamefont
  {Barsegov}(2002)}]{petrosky2002quantum}%
  \BibitemOpen
  \bibfield  {author} {\bibinfo {author} {\bibfnamefont {T.}~\bibnamefont
  {Petrosky}}\ and\ \bibinfo {author} {\bibfnamefont {V.}~\bibnamefont
  {Barsegov}},\ }\bibfield  {title} {\bibinfo {title} {{Quantum decoherence,
  Zeno process, and time symmetry breaking}},\ }\href@noop {} {\bibfield
  {journal} {\bibinfo  {journal} {Physical Review E}\ }\textbf {\bibinfo
  {volume} {65}},\ \bibinfo {pages} {046102} (\bibinfo {year}
  {2002})}\BibitemShut {NoStop}%
\bibitem [{\citenamefont {Garmon}\ \emph {et~al.}(2013)\citenamefont {Garmon},
  \citenamefont {Petrosky}, \citenamefont {Simine},\ and\ \citenamefont
  {Segal}}]{garmon2013amplification}%
  \BibitemOpen
  \bibfield  {author} {\bibinfo {author} {\bibfnamefont {S.}~\bibnamefont
  {Garmon}}, \bibinfo {author} {\bibfnamefont {T.}~\bibnamefont {Petrosky}},
  \bibinfo {author} {\bibfnamefont {L.}~\bibnamefont {Simine}},\ and\ \bibinfo
  {author} {\bibfnamefont {D.}~\bibnamefont {Segal}},\ }\bibfield  {title}
  {\bibinfo {title} {{Amplification of non-Markovian decay due to bound state
  absorption into continuum}},\ }\href@noop {} {\bibfield  {journal} {\bibinfo
  {journal} {Fortschritte der Physik}\ }\textbf {\bibinfo {volume} {61}},\
  \bibinfo {pages} {261} (\bibinfo {year} {2013})}\BibitemShut {NoStop}%
\bibitem [{\citenamefont {Cordero}\ and\ \citenamefont
  {Garc{\'\i}a-Calder{\'o}n}(2012)}]{cordero2012analytical}%
  \BibitemOpen
  \bibfield  {author} {\bibinfo {author} {\bibfnamefont {S.}~\bibnamefont
  {Cordero}}\ and\ \bibinfo {author} {\bibfnamefont {G.}~\bibnamefont
  {Garc{\'\i}a-Calder{\'o}n}},\ }\bibfield  {title} {\bibinfo {title}
  {{Analytical study of quadratic and nonquadratic short-time behavior of
  quantum decay}},\ }\href@noop {} {\bibfield  {journal} {\bibinfo  {journal}
  {Physical Review A—Atomic, Molecular, and Optical Physics}\ }\textbf
  {\bibinfo {volume} {86}},\ \bibinfo {pages} {062116} (\bibinfo {year}
  {2012})}\BibitemShut {NoStop}%
\bibitem [{\citenamefont {Muga}\ \emph {et~al.}(1996)\citenamefont {Muga},
  \citenamefont {Wei},\ and\ \citenamefont {Snider}}]{muga1996survival}%
  \BibitemOpen
  \bibfield  {author} {\bibinfo {author} {\bibfnamefont {J.}~\bibnamefont
  {Muga}}, \bibinfo {author} {\bibfnamefont {G.}~\bibnamefont {Wei}},\ and\
  \bibinfo {author} {\bibfnamefont {R.}~\bibnamefont {Snider}},\ }\bibfield
  {title} {\bibinfo {title} {{Survival probability for the Yamaguchi
  Potential}},\ }\href@noop {} {\bibfield  {journal} {\bibinfo  {journal}
  {Annals of Physics}\ }\textbf {\bibinfo {volume} {252}},\ \bibinfo {pages}
  {336} (\bibinfo {year} {1996})}\BibitemShut {NoStop}%
\bibitem [{\citenamefont {Li}\ \emph {et~al.}(2018)\citenamefont {Li},
  \citenamefont {Hall},\ and\ \citenamefont {Wiseman}}]{li2018concepts}%
  \BibitemOpen
  \bibfield  {author} {\bibinfo {author} {\bibfnamefont {L.}~\bibnamefont
  {Li}}, \bibinfo {author} {\bibfnamefont {M.~J.}\ \bibnamefont {Hall}},\ and\
  \bibinfo {author} {\bibfnamefont {H.~M.}\ \bibnamefont {Wiseman}},\
  }\bibfield  {title} {\bibinfo {title} {{Concepts of quantum non-Markovianity:
  A hierarchy}},\ }\href@noop {} {\bibfield  {journal} {\bibinfo  {journal}
  {Physics Reports}\ }\textbf {\bibinfo {volume} {759}},\ \bibinfo {pages} {1}
  (\bibinfo {year} {2018})}\BibitemShut {NoStop}%
\bibitem [{\citenamefont {Shrikant}\ and\ \citenamefont
  {Mandayam}(2023)}]{shrikant2023quantum}%
  \BibitemOpen
  \bibfield  {author} {\bibinfo {author} {\bibfnamefont {U.}~\bibnamefont
  {Shrikant}}\ and\ \bibinfo {author} {\bibfnamefont {P.}~\bibnamefont
  {Mandayam}},\ }\bibfield  {title} {\bibinfo {title} {{Quantum
  non-Markovianity: Overview and recent developments}},\ }\href@noop {}
  {\bibfield  {journal} {\bibinfo  {journal} {Frontiers in Quantum Science and
  Technology}\ }\textbf {\bibinfo {volume} {2}},\ \bibinfo {pages} {1134583}
  (\bibinfo {year} {2023})}\BibitemShut {NoStop}%
\bibitem [{\citenamefont {Seke}\ and\ \citenamefont
  {Herfort}(1988)}]{seke1988deviations}%
  \BibitemOpen
  \bibfield  {author} {\bibinfo {author} {\bibfnamefont {J.}~\bibnamefont
  {Seke}}\ and\ \bibinfo {author} {\bibfnamefont {W.}~\bibnamefont {Herfort}},\
  }\bibfield  {title} {\bibinfo {title} {{Deviations from exponential decay in
  the case of spontaneous emission from a two-level atom}},\ }\href@noop {}
  {\bibfield  {journal} {\bibinfo  {journal} {Physical Review A}\ }\textbf
  {\bibinfo {volume} {38}},\ \bibinfo {pages} {833} (\bibinfo {year}
  {1988})}\BibitemShut {NoStop}%
\bibitem [{\citenamefont {Berman}\ and\ \citenamefont
  {Ford}(2010)}]{berman2010spontaneous}%
  \BibitemOpen
  \bibfield  {author} {\bibinfo {author} {\bibfnamefont {P.~R.}\ \bibnamefont
  {Berman}}\ and\ \bibinfo {author} {\bibfnamefont {G.~W.}\ \bibnamefont
  {Ford}},\ }\bibfield  {title} {\bibinfo {title} {{Spontaneous decay,
  unitarity, and the Weisskopf--Wigner approximation}},\ }in\ \href@noop {}
  {\emph {\bibinfo {booktitle} {Advances in Atomic, Molecular, and Optical
  Physics}}},\ Vol.~\bibinfo {volume} {59}\ (\bibinfo  {publisher} {Elsevier},\
  \bibinfo {year} {2010})\ pp.\ \bibinfo {pages} {175--221}\BibitemShut
  {NoStop}%
\bibitem [{\citenamefont {Grimsmo}\ \emph {et~al.}(2013)\citenamefont
  {Grimsmo}, \citenamefont {Vaskinn}, \citenamefont {Rekdal},\ and\
  \citenamefont {Skagerstam}}]{grimsmo2013memory}%
  \BibitemOpen
  \bibfield  {author} {\bibinfo {author} {\bibfnamefont {A.~L.}\ \bibnamefont
  {Grimsmo}}, \bibinfo {author} {\bibfnamefont {A.~H.}\ \bibnamefont
  {Vaskinn}}, \bibinfo {author} {\bibfnamefont {P.~K.}\ \bibnamefont
  {Rekdal}},\ and\ \bibinfo {author} {\bibfnamefont {B.-S.~K.}\ \bibnamefont
  {Skagerstam}},\ }\bibfield  {title} {\bibinfo {title} {{Memory effects in
  spontaneous emission processes}},\ }\href@noop {} {\bibfield  {journal}
  {\bibinfo  {journal} {Physical Review A—Atomic, Molecular, and Optical
  Physics}\ }\textbf {\bibinfo {volume} {87}},\ \bibinfo {pages} {022101}
  (\bibinfo {year} {2013})}\BibitemShut {NoStop}%
\end{thebibliography}%

\endgroup

\end{document}